\documentclass[twocolumn]{aastex62}

\usepackage{color}
\newcommand{\clr}{black}

\accepted{May 7, 2019}
\submitjournal{ApJ}

\shorttitle{Hazes on GJ1214b}
\shortauthors{Lavvas et al.}

\begin{document}

\title{Photochemical hazes in sub-Neptunian atmospheres with focus on GJ 1214 b}

\correspondingauthor{Panayotis Lavvas}
\email{panayotis.lavvas@univ-reims.fr}

\author[0000-0002-5360-3660]{Panayotis Lavvas}
\affil{Groupe de Spectrom\'etrie Moleculaire et Atmosph\'erique, Universit\'e de Reims Champagne Ardenne, Reims, France}

\author{Tommi Koskinen}
\affil{Lunar and Planetary Laboratory, University of Arizona, Tucson, USA }

\author[0000-0001-8342-1895]{Maria Steinrueck}
\affil{Lunar and Planetary Laboratory, University of Arizona, Tucson, USA }

\author[0000-0003-1756-4825]{Antonio Garc\'ia Mu\~noz}
\affil{Zentrum f\"ur Astronomie und Astrophysik, Technische Universit\"at Berlin,
Berlin, Germany}

\author{Adam P. Showman }
\affil{Lunar and Planetary Laboratory, University of Arizona, Tucson, USA }

%
%
%
%



\begin{abstract}

We study the properties of photochemical hazes in super-Earths/mini-Neptunes atmospheres with particular focus on GJ1214b. We evaluate photochemical haze properties at different metallicities between solar and 10000$\times$solar. Within the four orders of magnitude change in metallicity, we find that the haze precursor mass fluxes change only by a factor of $\sim$3. This small diversity occurs with a non-monotonic manner among the different metallicity cases, reflecting the interaction of the main atmospheric gases with the radiation field. Comparison with relative haze yields at different metallicities from laboratory experiments reveals a qualitative similarity with our theoretical calculations and highlights the contributions of different gas precursors. Our haze simulations demonstrate that higher metallicity results into smaller average particle sizes. Metallicities at and above 100$\times$solar with haze formation yields of $\sim$10$\%$ provide enough haze opacity to satisfy transit observation at visible wavelengths and obscure sufficiently the H$_2$O molecular absorption features between 1.1 $\mu$m and 1.7 $\mu$m. However, only the highest metallicity case considered (10000$\times$solar) brings the simulated spectra into closer agreement with transit depths at 3.6 $\mu$m and 4.5 $\mu$m indicating a {\color{\clr}high} contribution of CO/CO$_2$ in GJ1214b's atmosphere. We also evaluate the impact of aggregate growth in our simulations, in contrast to spherical growth, and find that the two growth modes provide similar transit signatures (for D$_f$=2), but with different particle size distributions. Finally, we conclude that the simulated haze particles should have major implications for the atmospheric thermal structure and for the properties of condensation clouds.

~\\

\end{abstract}



\section{Introduction} \label{sec:intro}


Super-Earth and mini-Neptune size planets constitute a large fraction of the observed exoplanet population, therefore are the subject of intense investigation. Transit observations of such planets reveal a diversity in the depth of spectral features anticipated at near IR wavelengths from molecular absorption, which is attributed to the presence of suspended particulate matter in their atmospheres \citep{Crossfield17}. Such heterogeneous components have been detected also in the atmospheres of the larger hot-Jupiters \citep{Sing16,Barstow17}, for which various studies show that photochemical hazes and/or clouds are responsible for the observed signatures\footnote{We consider as hazes the end products of photochemical processes similar to those found in Titan's atmosphere or the soots of combustion/pyrolysis experiments, and as clouds the particles resulting from the condensation of gas phase molecules.}. Particularly for the case of HD 189733 b photochemical hazes appear to be better candidates for the interpretation of the observations \citep{ Lee15,Lavvas17,Powell18}. We explore here the properties of photochemical hazes in the atmospheres of super-Earths/mini-Neptunes using detailed models of atmospheric chemistry and haze microphysics, and evaluate how such components can help interpret the available observations. 

The most well studied example of such an atmosphere is the case of exoplanet GJ 1214 b, for which multiple ground based and space-borne observations reveal a remarkably flat spectrum extending from Visible to IR \citep[see][for an overview of available transit observations]{Angerhausen17}. A high precision evaluation of transit depth at visible wavelengths is prevented by the temporal and spatial inhomogeneities of the GJ 1214 M-dwarf stellar emission \citep{Rackham17, Mallonn18}. Nevertheless, the most accurate constraints at visible from ground based observations with the Large Binocular Telescope  \citep{Nascimbeni15}, in combination with the high-precision Hubble Space Telescope (HST) observations between 1.1 and 1.7 $\mu$m \citep{Kreidberg14}, and the Spitzer measurements at 3.6 and 4.5 $\mu$m \citep{Fraine13} reveal a quasi-flat spectrum, that restricts the characterisation of the atmospheric composition. Simulated transit spectra assuming only gaseous components are inconsistent with the observations even when high solar metallicities are assumed or even pure H$_2$O/CO$_2$ atmospheres, hence making the presence of the heterogeneous opacity necessary \citep{Kreidberg14}. Although the almost featureless transit spectrum of GJ 1214 b could be fitted by a planetary body with no substantial atmosphere, such a scenario seems unlikely as the measure radius and mass of the planet provide a bulk density  ($\rho$$\sim$2 g cm$^{-3}$) that is well below the typical rock/metal composition limit.
Mass and radius measurements cannot uniquely constrain the composition of the planet, however, thermal evolution models of the planet interior suggest that a few percent of the planet mass should be due to a H/He envelope to explain the measured mass density \citep{Lopez14,Lozovsky18}. Thus, the presence of an atmosphere on GJ 1214 b is rather probable.

Preliminary studies for the properties of clouds and photochemical hazes in the atmosphere of GJ 1214 b, suggested that both components can provide a flat spectrum \citep{Morley13, Morley15}. Photochemical hazes are expected to form in the upper atmosphere (p$<$10$\mu$bar) and coagulate to larger particles as they sediment to the lower atmosphere, while clouds are expected to form at {\color{\clr}higher} pressures depending on the solar metallicity and temperature conditions. However, the interpretation of the observations with condensates of KCl and ZnS composition (the most likely cloud condensation candidates for the anticipated atmospheric conditions) require high metallicity conditions (1000$\times$solar), as well as, significantly reduced sedimentation velocities for the formed cloud particles. On the contrary, photochemical hazes could explain the observed flatness of the HST observations at lower metallicities (50$\times$solar) and without the requirement of reduced sedimentation \citep{Morley15}.

Studies with general circulation models (GCM) suggest that atmospheric dynamics could uplift particles in the atmosphere of GJ 1214 b (at pressures higher than $\sim$0.1mbar), with increasing efficiency as the metallicity increases above solar \citep{Charnay15a}. Under the assumption that cloud particle sizes are constant with height a particle radius of 0.5 $\mu$m provided the best match to the HST observations \citep{Charnay15b}. 

Subsequent investigations focusing on the microphysics of cloud formation \citep{Gao18} demonstrate that KCl and ZnS clouds could match the available observations only at high metallicities (1000$\times$solar) and under the assumption of strong atmospheric mixing (K$_{ZZ}$=10$^{10}$cm$^2s^{-1}$). However, such an efficient mixing is not supported by the GCM results {\color{\clr}\citep[K$_{ZZ}$=10$^7$-10$^9$ cm$^2s^{-1}$ from][]{Charnay15a}}, while heterogeneous nucleation on the surface of photochemical hazes, not yet evaluated, could affect the formation of the condensates and modify the resulting cloud particle distribution \citep{Gao18}. A similar qualitative conclusion for the shortcomings of condensates to explain the GJ 1214 b transit observations was derived from \cite{Ohno18}. These results suggest that photochemical hazes could potentially provide a more realistic interpretation. 

Previous work on the microphysics of photochemical haze formation on sub-Neptune size planets demonstrated how the variable degree of haze production can result to different levels of transit spectra flattening \citep{Kawashima18} and how aggregate growth can modify the haze properties \citep{Adams19}. Here we explore further the details of the microphysical  description and evaluate the haze properties at different conditions of solar metallicity. We furthermore evaluate the differences between spherical and aggregate growth and discuss the possible implications of the formed hazes on the atmospheric structure. Our work addresses the general properties of photochemical hazes on super-Earth/mini-Neptunes atmospheres and for that reason we compare our results with laboratory experiments simulating the photochemical haze formation at such conditions \citep{He18, Horst18}.


\section{Atmospheric properties}



For the evaluation of the photochemical haze properties we need information for chemical composition of the atmosphere. We use the 1D non-equilibrium chemistry model we have developed for hot-Jupiters \citep{Lavvas14} for the investigation of the chemical composition of GJ 1214 b. The required inputs for these simulations are the main atmospheric composition, the atmospheric thermal structure and mixing magnitude (K$_{ZZ}$), and the stellar insolation. 

\begin{figure}[!t]
\centering
\includegraphics[scale=0.47]{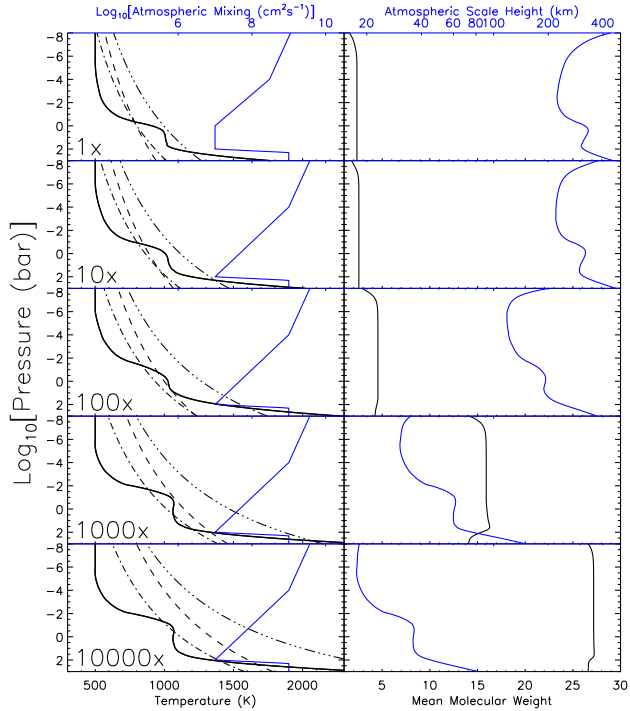}
\caption{Atmospheric structure assumed for GJ 1214 b at different metallicities. Left side panels present the thermal structure (black solid lines) based on \cite{Gao18}. The broken lines present the condensation curves for ZnS (dashed), KCl (dash-dotted), and Na$_2$S (dash-triple-dotted) assuming thermochemical equilibrium \citep{Morley12}, while the blue curves present the assumed atmospheric mixing profile for each case using input from the GCM models \citep{Charnay15a}. The right side panels present the mean molecular weight (black) and scale height (blue) of the atmosphere based on our disequilibrium chemical composition.}\label{temperature}
\end{figure}

The main composition of GJ 1214 b is unknown and a topic of debate \citep{Kreidberg14, Charnay15b, Marley13}. Cases between solar metallicity to H$_2$O or CO$_2$ dominated atmospheres have been studied previously, but no definite conclusion could be derived from the available observations due to the implications of the heterogeneous opacity contribution. As the formation of hazes or clouds depends on  species other than the main composition, we need to {\color{\clr}resort} to the approach of a metallicity scaling factor, i.e. evaluate the chemical composition at different scalings of the solar metallicity. Previous studies indicate that a high (1000$\times$solar) metallicity is required to explain the transit spectrum of GJ 1214 b, if KCl and ZnS clouds are assumed the source of the heterogeneous opacity \citep{Gao18}. However, an explanation based on photochemical aerosols could be less constraining on the metallicity factor. Laboratory studies on the production of photochemical aerosols on super-Earths and mini-Neptunes demonstrate a large diversity in the production rates \citep{Horst18}. Thus, we proceed in estimating photochemical haze production rates at 1$\times$, 10$\times$, 100$\times$, 1000$\times$, and 10000$\times$solar metallicity and discuss how these correlate with the available observational constraints.

For the thermal structure we consider the available profiles reported in the literature for the different metallicity cases we consider \citep{Gao18}, which are based on 1D thermal structure calculations assuming thermochemical equilibrium. For the 10000$\times$solar metallicity case we assume the same profile as for the 1000$\times$ case based on previous estimates \citep{Moses13}. General circulation models also provide significant information about how these profiles vary around the planet and particularly for the atmospheric regions probed during transit observations \citep{Kataria14, Charnay15a}. The disk average temperature profiles provided by GCMs are consistent with the 1D calculations, moreover they demonstrate that center-to-limb temperature changes could be up to $\pm$100 K above 10 mbar. \cite{Morley15} and \cite{Charnay15b} demonstrated that heterogeneous opacities of both clouds or hazes can have an impact on the atmospheric thermal structure, resulting in an anti-greenhouse effect that forces stellar visible radiation to be deposited at higher altitudes relative to a clear atmosphere. However, higher energy photons leading to the haze formation are deposited at much higher altitudes, while, although the detailed pathways of haze formation are not clear at the moment, temperature changes are expected to have a smaller impact on haze properties relative to cloud properties, as long as, the temperature remains below the thermal decomposition limit of the haze particles \citep{Lavvas17}. Photochemical hazes form from a multitude of ion and neutral chemical processes each with its individual temperature dependence, while condensation depends dominantly on the saturation vapour pressure of the condensing component that has a sharp temperature dependence \citep{Morley12}. Therefore, we do not consider the implications of temperature changes in our current calculations. Finally, we assume isothermal profiles in the upper atmosphere although thermospheres are anticipated to form \citep{GarciaMunoz07,Koskinen13}. However, the chemical complexity anticipated at high metallicity cases is likely to significantly modify the current understanding of thermospheres based on H$_2$/H/He atmospheres of hot-Jupiters. Thus, a correct evaluation of the thermospheric structure requires an investigation of the coupling between the lower and upper atmosphere over different metallicities and will be addressed separately.

\begin{figure}[!t]
\includegraphics[scale=0.5]{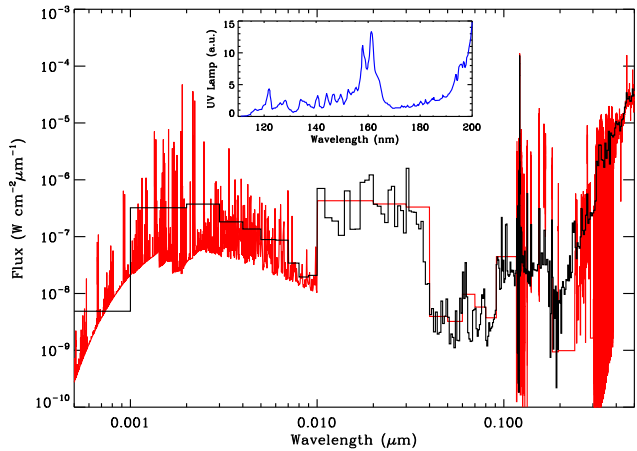}
\caption{Stellar emission from GJ 1214 assumed in the calculations. The red line shows the high-resolution spectrum based on the MUSCLES database, and the black line depicts the spectrum mapped to the resolution of our model (see text). The inset presents the UV light lamp used for laboratory haze formation for comparison \citep{He18}. } \label{star}
\end{figure}

For the 1D atmospheric mixing profile we use results from general circulation models at different metallicities \citep{Charnay15a}. According to these results the atmospheric mixing is similar for metallicities between 10x and 100x solar with the mixing magnitude increasing from 10$^7$ cm$^2$s$^{-1}$ at 100 bar to 10$^9$ cm$^2$s$^{-1}$ at 0.1 mbar. For the 1x solar case the mixing profile is weaker by about a factor of 10 relative to the higher metallicity cases. It is well acknowledged that the eddy mixing magnitude derived by circulation models can only be treated as approximate estimates \citep{Parmentier13, Charnay15a} subject to uncertainties possibly greater than a factor of 10. Thus, we treat the assumed K$_{ZZ}$ profiles as approximate. 

For the stellar flux of GJ 1214 (M4.5, T$_{eff}$ = 2935 K, log $g$(cms$^{-2}$) = 5.06) we use input from the MUSCLES database (Fig.~\ref{star}) that combines observations with models of stellar emissions \citep{France16, Loyd16}. Observations with HST/STIS allow for the reconstruction of the Ly-$\alpha$ emission and provide a flux of 1.3$\times$10$^{-14}$ erg cm$^{-2}$ s$^{-1}$ \AA$^{-1}$ reaching the Earth \citep{Youngblood16} . This value can be used to estimate the EUV emission at broad bins between 100 \AA ~and 1170 \AA ~\citep{Linsky14}. The spectral density within each EUV bin is unknown, thus we assume a solar-type spectral distribution. A similar combination of X-ray observations and simulations provides the spectrum at shorter wavelenghts \citep{Fontenla16}. GJ 1214 is at 14.6 pc, which is sufficiently distant that there exist large uncertainties in the observed spectrum. However, observations of a similar spectral type star, GJ 876 (M5, T$_{eff}$ = 3062 K, log $g$(cms$^{-2}$) = 4.93) at 4.7 pc provide an improved accuracy spectrum. Thus, we used the spectral density of GJ 876 to estimate the spectral emission of GJ 1214 for regions where noise dominated the observations, particularly at the Ly-$\alpha$ wings and up to $\sim$3000\AA. At longer wavelengths the MUSCLES spectrum is based on observations with HST/STIS and HST/COS combined with PHOENIX stellar emission models. 

At the lower boundary {\color{\clr}(10$^3$ bar)} we consider abundances derived from the CEA thermochemical equilibrium model \citep{McBride02}. The various metallicity cases we consider result in different main atmospheric compositions that reflect on the mean molecular weight and the corresponding atmospheric viscosity. These parameters are important for the simulation of molecular diffusion and settling velocities of haze particles and are calculated self-consistently as a function of pressure within our model. The mean molecular weight is calculated from our chemical composition calculations, while for the atmospheric viscosity we use the  corresponding state method for gas mixtures, taking into account the impact of pressure on the gas viscosity {\color{\clr}at p$>$}10 bar \citep{Poling01}.

\begin{table}
\caption{Main atmospheric composition at 1 mbar for different metallicity cases according to the disequilibrium chemistry calculations. Read a(b) as a$\times$10$^b$.}\label{comp1mbar}
\begin{tabular}{lccccc}
\hline
Species 	& 1x 		&10x 	& 100x 	& 1000x	& 10000x \\
\hline
H$_2$	& 0.835	& 0.822	& 0.706	& 0.341	& 7.08(-2)\\
He		& 0.163	& 0.164	& 0.167	& 0.111	& 2.14(-2)\\
H$_2$O	& 7.13(-4)	& 7.12(-3)	& 5.58(-2)	& 7.77(-2)	& 1.22(-2)\\
CH$_4$	& 5.94(-4)	& 6.03(-3)	& 4.64(-2)	& 6.74(-2)	& 1.05(-2)\\
N$_2$	& 3.37(-5)	& 7.79(-4)	& 9.52(-3)	& 6.39(-2)	& 0.123 \\
NH$_3$	& 1.15(-4)	& 3.26(-4)	& 3.66(-4)	& 4.10(-4)	& 5.39(-5)\\
CO		& 5.16(-7)	& 1.56(-5)	& 1.35(-2)	& 0.274	& 0.651\\
CO$_2$	&2.11(-10)& 2.16(-7)	& 1.61(-3)	& 6.43(-2)	& 0.111\\
\hline
\end{tabular}
\end{table}

\begin{figure*}[!ht]
\centering
\includegraphics[scale=0.5]{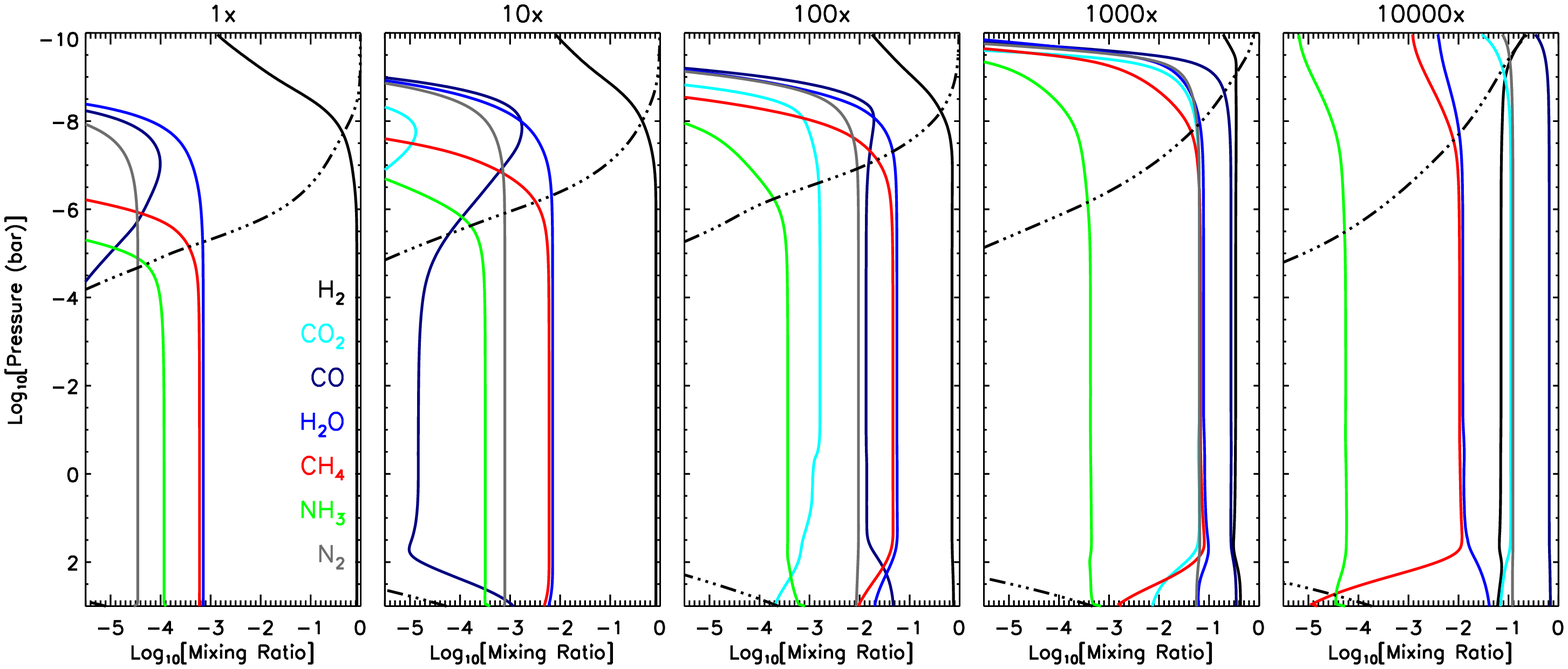}
\includegraphics[scale=0.5]{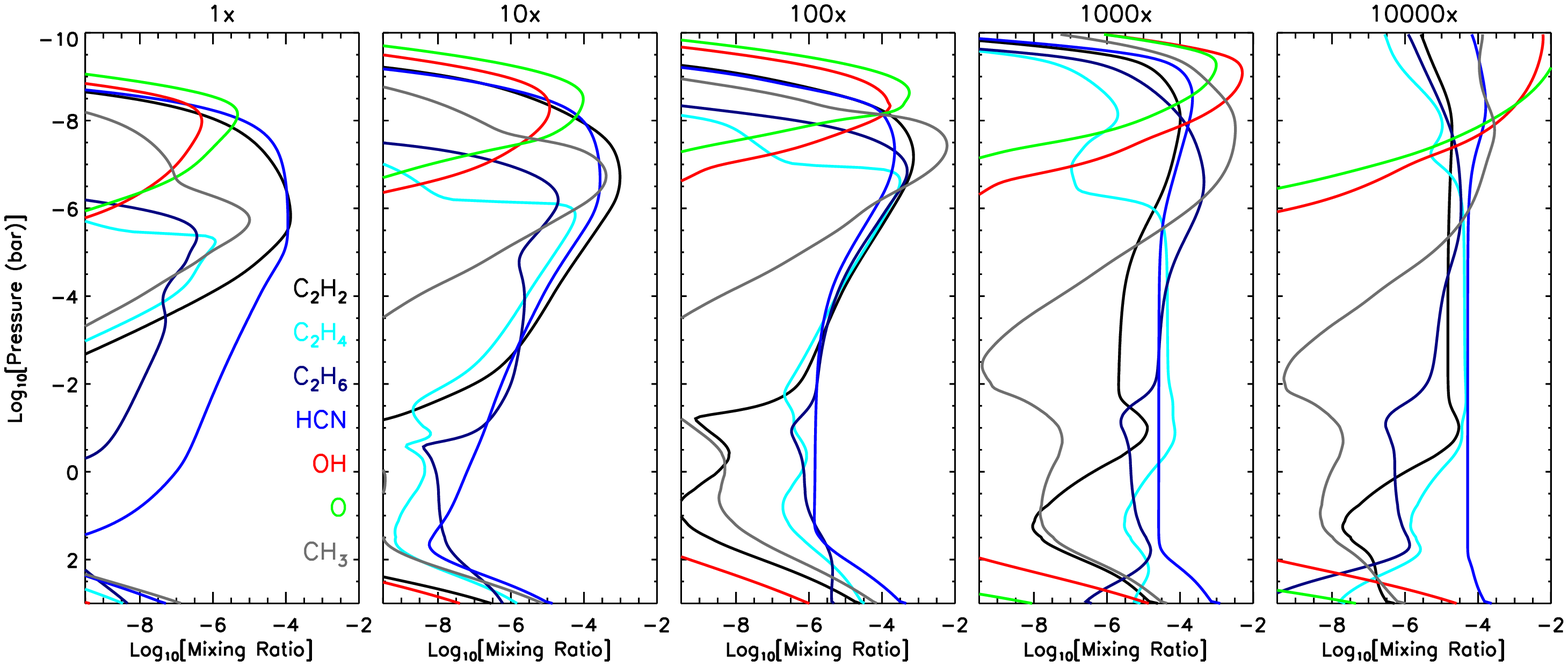}
\caption{Chemical composition of GJ 1214 b for the five metallicity cases and for disk geometry. The top panels present the variation of the main composition (broken line corresponds to atomic hydrogen) and the bottom panels the corresponding photochemical products of hydrocarbons and oxygen species.}\label{composition}
\end{figure*}

\section{Haze production}

With the above inputs we calculate the chemical composition of GJ 1214 b under different metallicity assumptions (Fig.~\ref{composition}). Increasing the atmospheric metallicity leads to an increase in the atmospheric mean molecular weigh with a concurrent decrease in the atmospheric scale height (Fig.~\ref{temperature}). The mixing ratios of main atmospheric components in the middle atmosphere (1 mbar) are shown in Table~\ref{comp1mbar} for each metallicity case. The chemistry of hot-Neptune atmospheres was investigated in detail in previous studies \citep[e.g.][]{MillerRicciKempton12,Moses13}, thus we will not dwell on these details here. As these studies demonstrated, as metallicity increases the abundance of CO/CO$_2$ progressively dominates over the CH$_4$ abundance and quenching in the deep atmosphere from atmospheric mixing will affect the relative abundance of CH$_4$/CO in the upper atmosphere. We consider a nominal values of  K$_{\rm ZZ}$=$10^9$ cm$^{2}$s$^{-1}$ for the convective region below 100 bar based on \cite{Moses13}. However, a smaller magnitude K$_{ZZ}$=$10^7$ cm$^{2}$s$^{-1}$ did not provide different results for all metallicity cases considered. 

\begin{table*}
\caption{Photolysis mass fluxes (g cm$^{-2}$s$^{-1}$) in the upper atmosphere of GJ 1214 b under different assumptions of solar metallicity. Soot mass fluxes are based on the contributions of CH$_4$, NH$_3$, HCN, C$_2$H$_2$, C$_2$H$_4$, and N$_2$ scaled by a yield of 1$\%$.}\label{Mfluxes}
\centering
\begin{tabular}{lccccc|ccccc}
\hline
			& \multicolumn{5}{c}{Metallicity - Limb}  & \multicolumn{4}{c}{Metallicity - Disk } \\
\hline
Species		&1$\times$& 10$\times$&100$\times$ &1000$\times$ & 10000$\times$&	1$\times$	& 10$\times$	&100$\times$ 	&1000$\times$&10000$\times$ 	\\
\hline
CH$_4$		&9.1(-14)	&1.1(-12)	&2.6(-12)	&7.4(-12)	& 2.9(-12)		&1.5(-13) & 4.8(-13) & 8.9(-12) & 7.4(-11) & 3.5(-11)\\
NH$_3$		&3.9(-12)	&3.9(-12)	&2.8(-12)	&1.7(-12)	& 9.3(-13)		&1.3(-11) & 2.6(-11) & 4.0(-11) & 3.4(-11) & 6.3(-12)\\
HCN			&2.9(-12)	&1.0(-12)	&1.1(-13)	&8.7(-14)	& 5.4(-14)		&1.6(-11) & 6.8(-12) & 1.7(-12) & 1.9(-12) & 2.3(-12) \\
C$_2$H$_2$	&1.2(-11)	&3.2(-12)	&3.4(-13)	&8.9(-14)	& 9.1(-14)		&1.0(-10) & 8.9(-11) & 1.5(-11) & 1.3(-12) & 5.2(-13) \\
C$_2$H$_4$	&4.6(-16)	&2.8(-15)	&3.1(-15)	&1.5(-14)	& 1.1(-13)		&2.6(-13) & 1.1(-12) & 6.2(-12) & 4.8(-13) & 1.0(-12) \\
N$_2$		&1.7(-32)	&3.2(-25)	&2.6(-23)	&5.7(-17)	& 5.8(-14)		&2.0(-20) & 6.5(-17) & 6.8(-16) & 3.8(-14) & 1.8(-12) \\
CO$_2$		&5.8(-16)	&4.7(-15)	&6.3(-14)	&1.3(-12)	& 2.9(-12)		&6.2(-15) & 1.0(-13) & 2.0(-12) & 3.9(-11) & 6.9(-11)\\
CO			&3.7(-24)	&3.7(-19)	&1.2(-17)	&1.5(-14)	& 1.0(-12)		&6.9(-17) & 2.4(-14) & 2.4(-13) & 2.8(-12) & 3.1(-11)\\
H$_2$O		&2.6(-11)	&2.6(-11)	&1.8(-11)	&1.4(-11)	& 3.7(-12)		&1.5(-10) & 1.5(-10) & 1.7(-10) & 1.8(-10) & 5.2(-11) \\
\hline
Soot(1$\%$)	&1.9(-13)	&9.2(-14)	&5.9(-14)	&9.3(-14)	& 1.2(-13)		&1.3(-12) & 1.2(-12) &7.2(-13) & 1.1(-12) & 4.7(-13)\\
\hline
\hline
\end{tabular}
\end{table*}

Our focus here is to evaluate the production rate of soot-type composition photochemical hazes for each case. As in our previous study on hot-Jupiters \citep{Lavvas17}, we estimate the haze mass flux from the photolysis mass fluxes generated by the photodissociation/photoionization of major species  in the upper atmosphere (p$<$10$\mu$bar) multiplied with an approximate yield of haze formation. We consider contributions from the photolysis of CH$_4$ and N$_2$ and from their main photochemical products HCN, C$_2$H$_2$, C$_2$H$_4$, as well as, from NH$_3$ (see Table~\ref{Mfluxes}). A similar approach for the evaluation of photochemical haze mass fluxes was followed in previous studies \citep{Morley13,Kawashima18}, although differences exist on the details of the calculation with other studies focusing on the abundance of various photochemical products and not their photolysis rates. 

Hazes based on sulfur composition are also a possible candidate for the temperature conditions of mini-Neptunes/super-Earths \citep{Zahnle16,Gao17}. However, for the solar elemental compositions we consider C is more abundant than S, thus is likely to provide a higher abundance of soot type hazes. Moreover, a correct representation of sulfur components requires consideration of cloud formation as ZnS and Na$_2$S are likely condensates at the conditions under investigation that will limit the abundance of sulfur-based haze precursors reaching the upper atmosphere. Thus, we focus here on the formation of soot-type haze precursors.

The haze formation yield for HD 189733 b was found to range between 0.5$\%$ and 10$\%$ depending on the atmospheric conditions assumed \citep{Lavvas17}, while we know that Titan's atmosphere (the most hazy in the solar system) has a haze yield of $\sim$30$\%$ \citep{Lavvas11,Lavvas13}. Thus, for our preliminary estimates of haze mass fluxes we consider a conservative 1$\%$ yield in our tabulated values (see Table~\ref{Mfluxes}). We use this estimate only as a tool for comparison among the different metallicity cases, as we evaluate further below what would be the required yield to explain the transit observations. We present chemical composition calculations for both limb (calculated under spherical geometry for tangent rays) and disc (dayside stellar flux for incidence angle of $\mu_0$=0.5) geometry conditions to evaluate what would be the difference in the haze production rate at different locations (given the relatively small, $\pm$100 K, horizontal temperature difference suggested by the GCM we consider the same temperature profile for different geometries). However, we assume that the atmospheric circulation will act towards a homogenisation of the particle distribution, although this is something that needs to be evaluated with a coupled model of atmospheric circulation and haze microphysics. Nevertheless, our simulations suggest that significant mass fluxes of photochemical hazes can exist both at limb and disk locations. 

\begin{figure}
\centering
\includegraphics[scale=0.5]{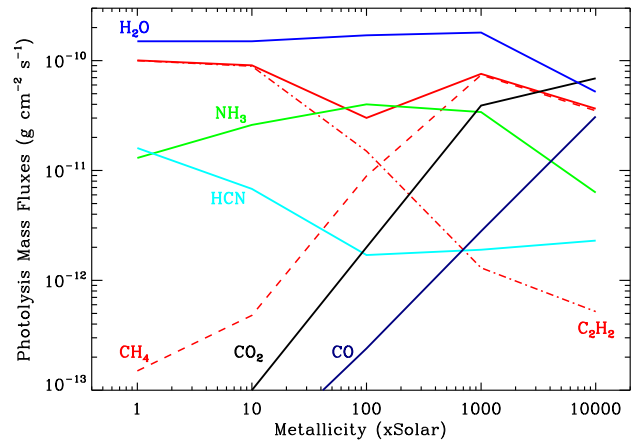}
\caption{Photolysis mass fluxes from various molecules for different metallicity cases.}\label{fluxes}
\end{figure}

With the assumed 1$\%$ soot formation yield, haze mass fluxes range between $\sim$2$\times$10$^{-14}$ gcm$^{2}$s$^{-1}$ and $\sim$10$^{-13}$ gcm$^{2}$s$^{-1}$ for limb geometry and between 7$\times$10$^{-13}$ gcm$^{2}$s$^{-1}$ and 1.3$\times$10$^{-12}$ gcm$^{2}$s$^{-1}$ for disk geometry, for the different metallicity cases. However, increasing the atmospheric metallicity causes changes in the haze production that are not monotonic (Table~\ref{Mfluxes}). Using the 100$\times$solar metallicity results as a reference case we find that lower metallicity conditions result in higher haze mass fluxes by factors of 1.8 and 1.7 at 1$\times$ and 10$\times$ solar metallicity, respectively. At 1000$\times$solar metallicity the haze production increases again by a factor of 1.5 before dropping by a factor of 0.65 at 10000$\times$. The reasons for these variations are due to the changes in the main atmospheric composition at each metallicity case and their implications on the photolysis rates.

\begin{figure*}
\centering
\includegraphics[scale=0.5]{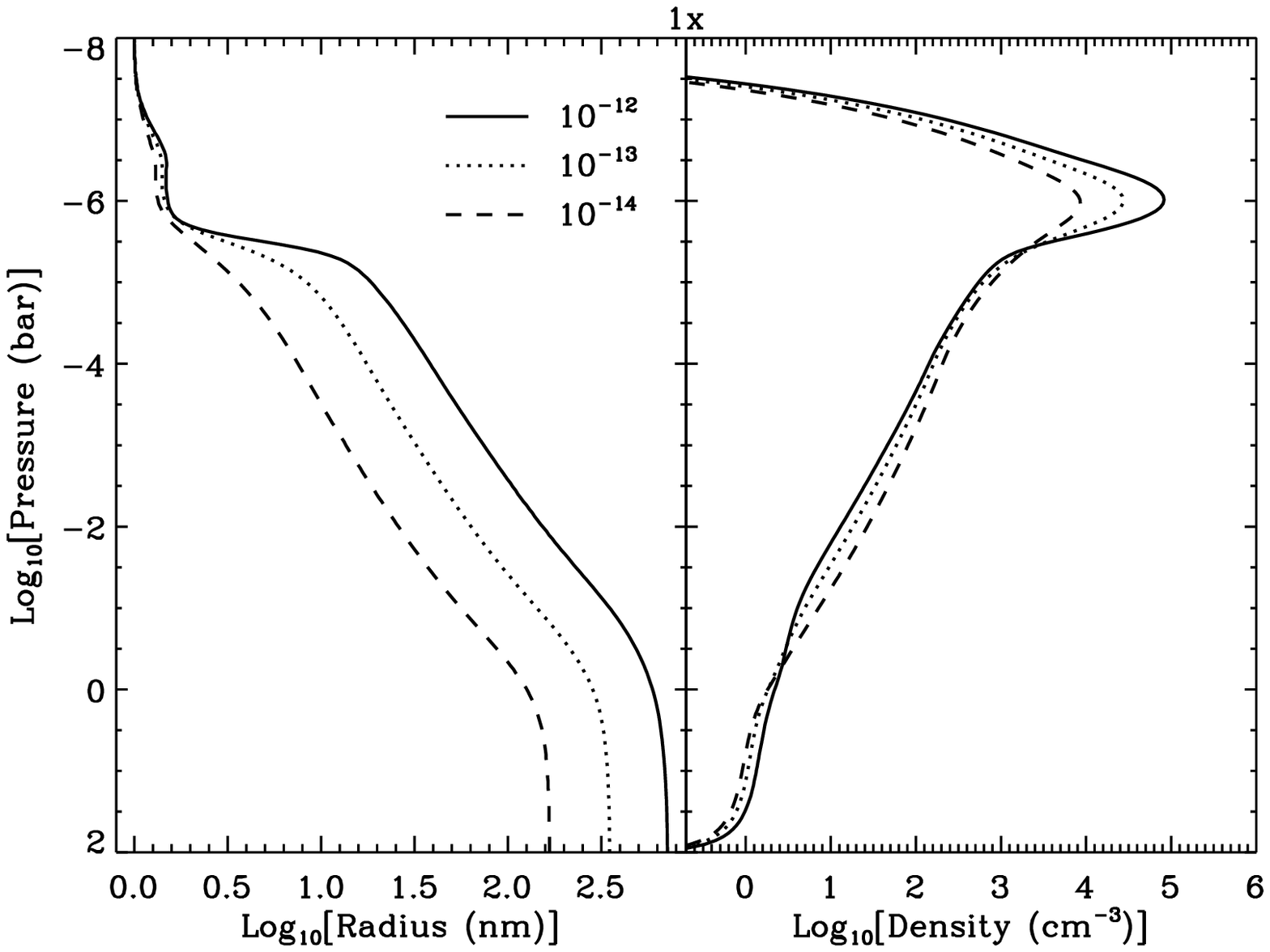}
\includegraphics[scale=0.5]{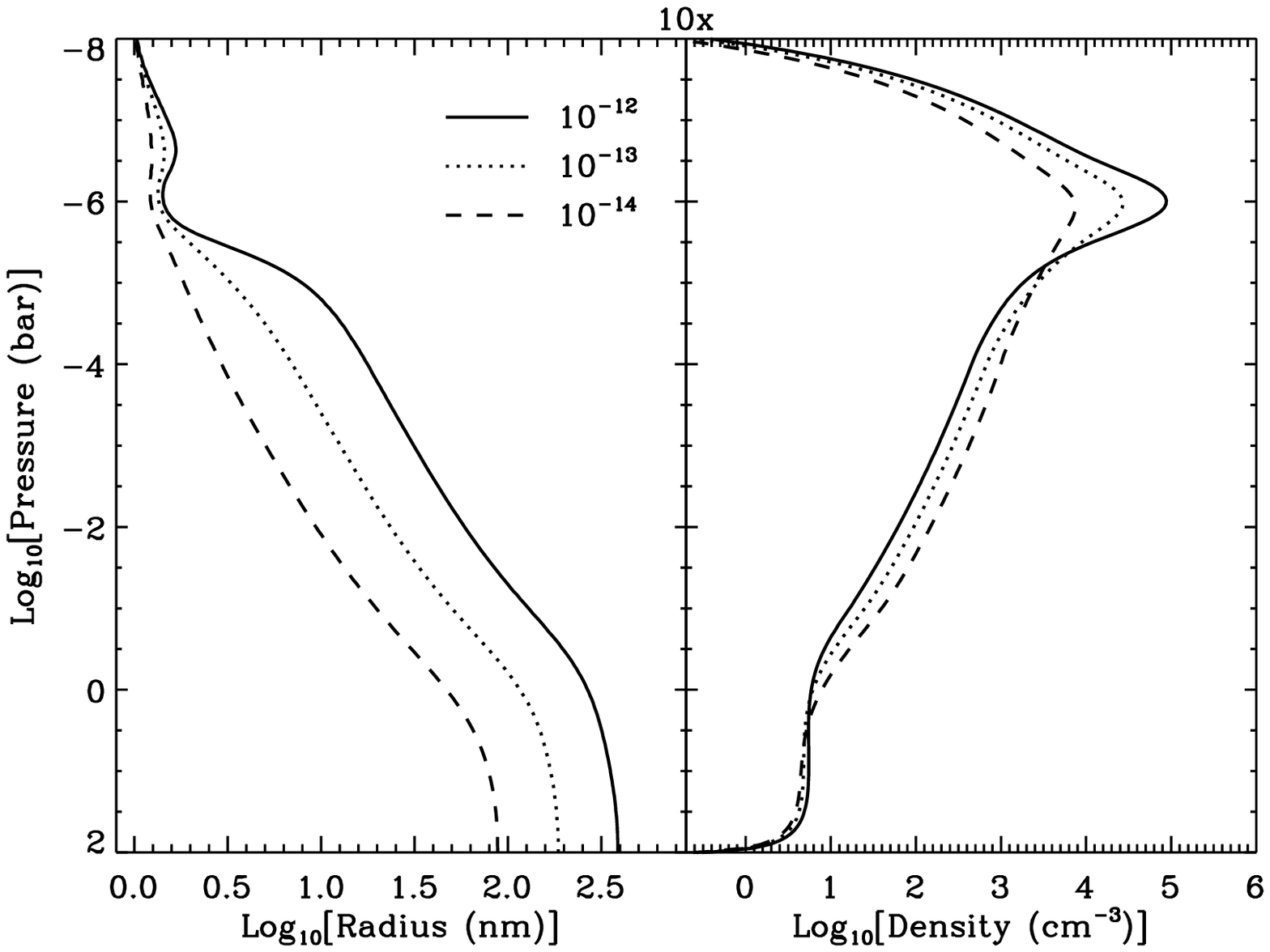}
\includegraphics[scale=0.5]{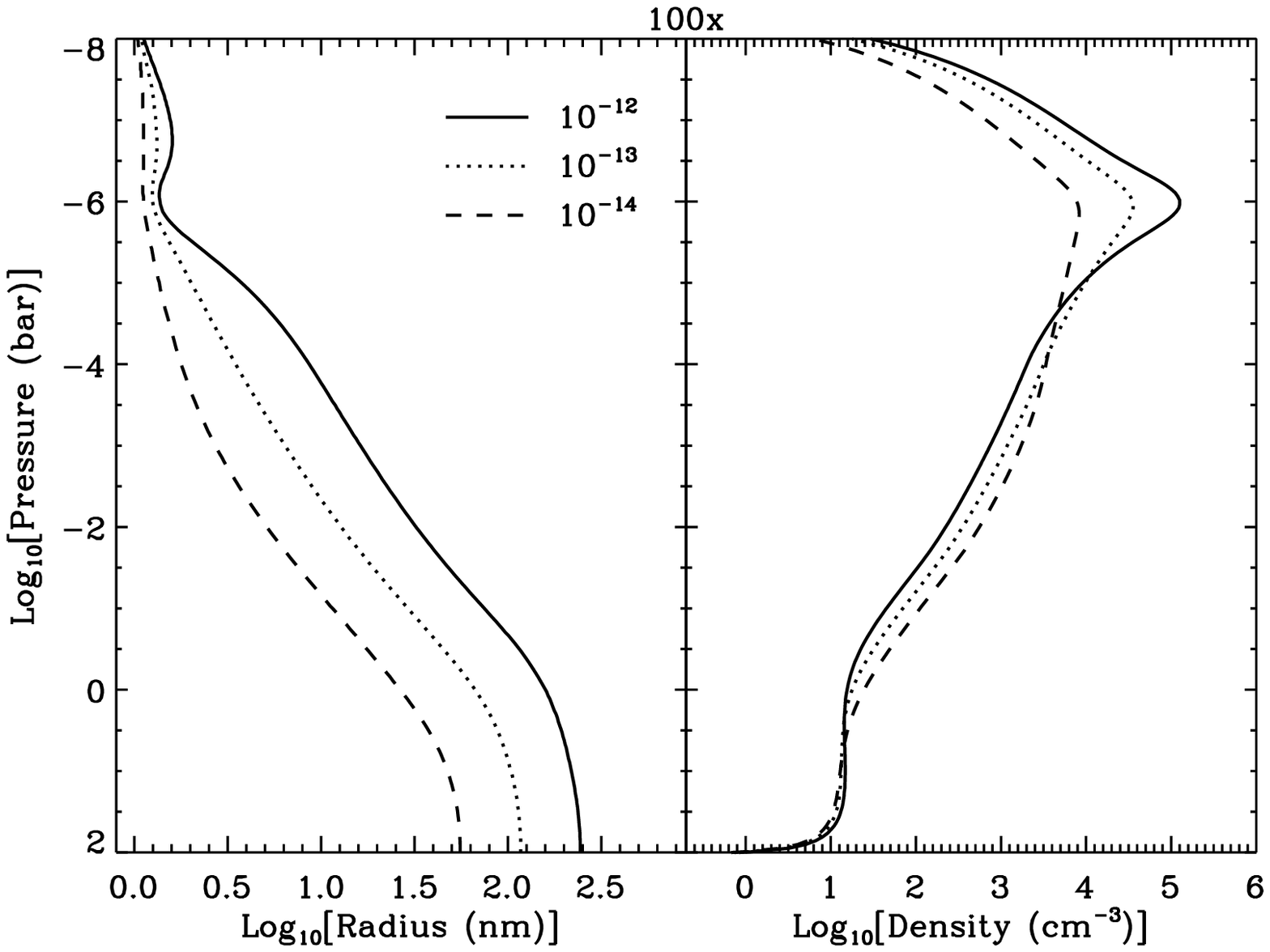}
\includegraphics[scale=0.5]{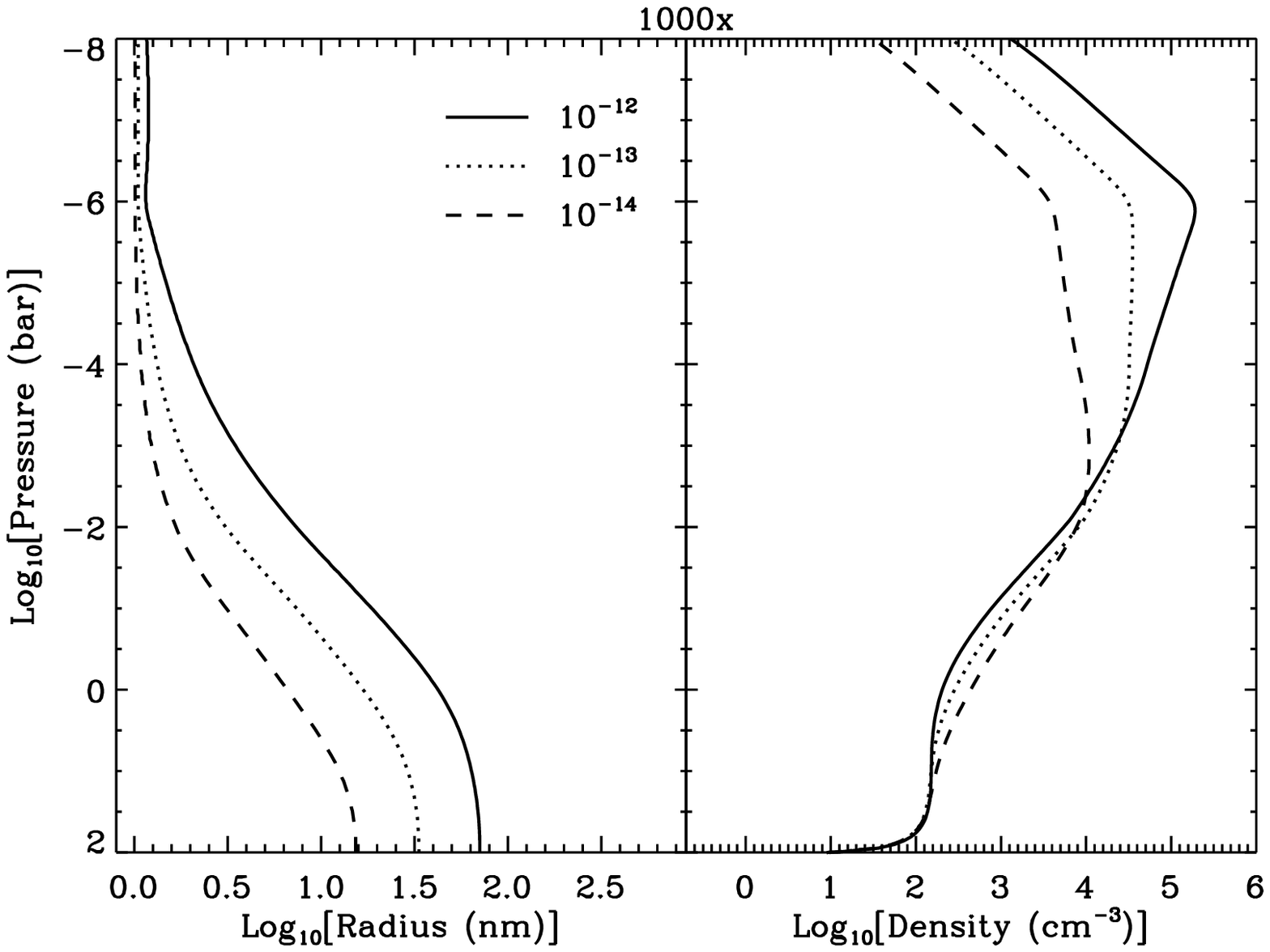}
\caption{Average radius and corresponding number density of haze particles in the atmosphere of GJ 1214 b, under different assumptions of haze mass flux and metallicity. }\label{micro}
\end{figure*}

At low metallicities the catalytic destruction of H$_2$ through interaction with OH, produced in the photodissociation of H$_2$O, leads to the formation of a large abundance of atomic hydrogen in the upper atmosphere \citep{Moses11}. The latter is a major pathway for the breakup of CH$_4$ (H + CH$_4$ $\rightarrow$ H$_2$ + CH$_3$) and the formation of higher order hydrocarbons. Therefore, for the solar metallicity case, at pressures lower than 1 $\mu$bar methane is lost and C$_2$-type hydrocarbons are the main hydrocarbon precursors for photochemical haze formation (Fig.~\ref{composition}). As metallicity increases to 10$\times$ and 100$\times$ solar, the atmospheric pressure where the catalytic destruction of H$_2$ takes place moves to lower pressures because the abundance of H$_2$O increases and UV photons are consumed more efficiently in the upper atmosphere leading to smaller penetration in the atmosphere. Therefore, more methane survives in the upper atmosphere and the production of higher order hydrocarbons is reduced. Hence, the photolysis contributions of higher order hydrocarbons decrease and that of CH$_4$ increases (Table~\ref{Mfluxes} $\&$ Fig.~\ref{fluxes}). Moreover, as metallicity increases, UV photons leading to H$_2$O photolysis are absorbed by other species as well, leading effectively to a reduction of the atomic hydrogen abundance. This reduction of atomic hydrogen in the upper atmosphere is more prominent for the 1000$\times$ and 10000$\times$ solar metallicity cases (Fig.~\ref{composition}). At these conditions, the relative contribution of direct photolysis of CH$_4$ becomes more important and provides the local increase in the haze production rate in the 1000$\times$solar metallicity case. At 10000$\times$solar metallicity, the CH$_4$ abundance is reduced, while the opacity of other competing species such as CO and CO$_2$ is higher leading to a reduction in methane photolysis (as well as that of other haze precursor species) and an overall decrease of haze production (Fig.~\ref{fluxes}). Below, we further discuss these results in the context of laboratory experiments. 








\section{Haze properties}

\begin{figure*}
\centering
\includegraphics[scale=0.5]{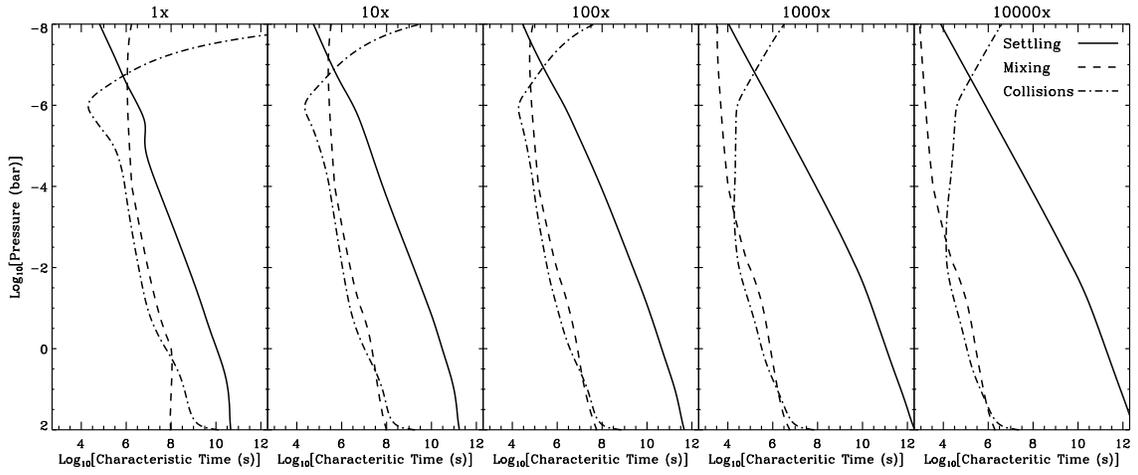}
\caption{Characteristic times for settling, mixing, and coagulation (collisions) for the haze particles of GJ 1214 b's atmosphere, for different metallicity cases. The haze mass flux is 10$^{-13}$ gcm$^{-2}$s$^{-1}$.}\label{times}
\end{figure*}

We can now use the above estimated haze fluxes to simulate the haze particle properties in the atmosphere of GJ 1214 b using the haze microphysics model developed for hot-Jupiters \citep{Lavvas17}. In a nutshell, the model solves for the evolution in size (radius) and number density of an initial population of haze embryos, utilising a bin of particle sizes and solving the continuity equation for each size particle, with contributions from atmospheric mixing, particle sedimentation, and particle coagulation. The initial particle production is described through a gaussian vertical production profile centred at 1 $\mu$bar with a radius of 1 nm. In order to convert the haze mass flux to a production rate of particles, we consider a particle mass density of 1 g cm$^{-3}$, a typical value assumed for planetary hazes \citep{Lavvas10,Rannou10} consistent also with laboratory experiments \citep{Horst13}. At the lower boundary particles are lost at the rate they arrive. 

Particles grow through their mutual collisions (coagulation). There are multiple mechanisms that can affect the particle collision rate (coagulation kernels) such as the particle brownian motion, gravitational settling, turbulent motions and convection \citep{Pruppacher78}. We have tested all these contributions and found that brownian motion is the dominant mechanism under the conditions we explore. This conclusion depends on the assumptions made for the particle production profile. For example, assuming that haze particles can form everywhere in the atmosphere would result in a bimodal distribution with small particles locally produced by the chemistry and a larger size component that originates from the sedimentation of particles coagulated at higher altitudes. Such a configuration would enhance contributions from the gravitational settling component of the coagulation kernel because of the different sedimentation velocities of particles found in the two main peaks of the bimodal distribution \citep{Kawashima18}. 

In our simulations we assume that production occurs only in the upper atmosphere, thus the shape of the particle distribution is mono-modal and the gravitational kernel has a minor contribution relative to brownian motion. Our choice for the particle production profile is based on the current understanding of haze formation in Titan's atmosphere \citep{Lavvas11a, Lavvas13}. There, haze formation initiates in the upper atmosphere through ion-neutral processes and subsequent growth of formed particles proceeds through heterogeneous processes on their surface. Thus, chemical processes occurring at lower altitudes do not result in the inception of new particles but contribute to the growth of the existing particles. Of course bimodal particle distributions do exist in planetary atmospheres but are related to different formation mechanisms or dynamical effects from the atmospheric circulation. Therefore, as we consider only a single type of haze in our simulations, we utilise a production profile located in the upper atmosphere. 


Our simulations reveal a drastic change in the haze particle properties among the different metallicity cases considered (Fig.~\ref{micro}). As metallicity increases the average particle size below the production region decreases, with the corresponding particle number density increasing. This feature is common for all mass flux cases considered. To interpret this effect we need to look into the processes defining the particle growth. 
As discussed above, changing the atmospheric metallicity results to changes in the mean molecular weight and the viscosity of the atmosphere, both increasing with the degree of metallicity. A higher viscosity decreases the particle settling velocity \citep{Lavvas10}, which would suggest that particle growth should increase with metallicity, as particles have a larger residence time in the atmosphere and are subject to more collisions. Similarly, the coagulation rates increase as the metallicity increases according to our calculations, due to the higher temperature. Moreover, the atmospheric mixing profile (eddy) is similar among the 3 highest metallicity cases implying that its role should be similar. However, the differences in the atmospheric properties among the various cases changes the response time for each of the three above processes (settling, coagulation, mixing), and result in the counter intuitive drop of particle growth with metallicity. The change in the characteristic times for particle settling, $\tau_S$, particle mixing, $\tau_M$,  and particle collisions, $\tau_C$, among the different atmospheric conditions highlight this effect (Fig.~\ref{times}). These times are calculated from:
\begin{equation}
\tau_{S} = \frac{H}{<v_p>}, ~~\tau_{M} = \frac{H^2}{K_{ZZ}}, ~~ \tau_{C} = <\frac{n_p}{\partial n_p/\partial t}>  
\end{equation}
with brackets corresponding to averaging over the particle size distribution. 
At solar metallicity, particle collisions have the smallest characteristic time leading to a rapid growth of the average particle size. Only below $\sim$1bar, atmospheric mixing due to the strong inversion assumed in the eddy profile, dominates over collisions and the particles are rapidly transferred towards the lower boundary, thus explaining the rapid drop on the particle number density observed for all metallicity cases at that location. At higher metallicities, as the mean molecular weight increases and the atmospheric scale-height decreases, the characteristic time for atmospheric mixing in the upper atmosphere decreases. Thus even if the eddy mixing profile is the same for the 10$\times$, 100$\times$, and 1000$\times$ solar metallicity cases, the characteristic time of atmospheric mixing becomes progressively smaller and particles are subject to less collisions. Thus, the rate of increase in the average particle size becomes progressively smaller with increasing metallicity (Fig.~\ref{micro}). Our simulations show that at 1 mbar and for a haze mass flux of 10$^{-12}$ gcm$^{-2}$s$^{-1}$ the average particle size decreases from $\sim$75 nm to $\sim$3 nm between the 1$\times$ and the 10000$\times$solar metallicity cases, while the corresponding particle number densitiy increases from $\sim$50 cm$^{-3}$ to $\sim$7$\times$10$^{4}$ cm$^{-3}$. 

\begin{figure*}[!t]
\centering
\includegraphics[scale=0.35]{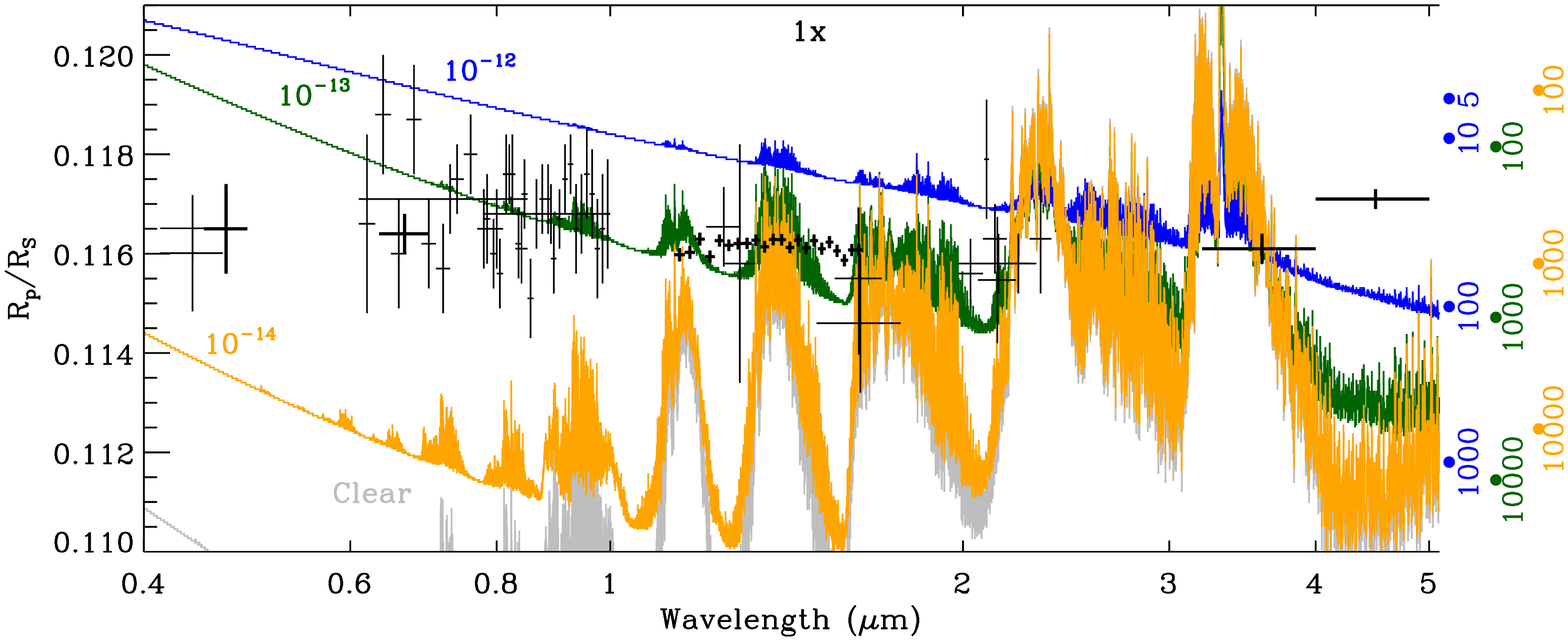}
\includegraphics[scale=0.35]{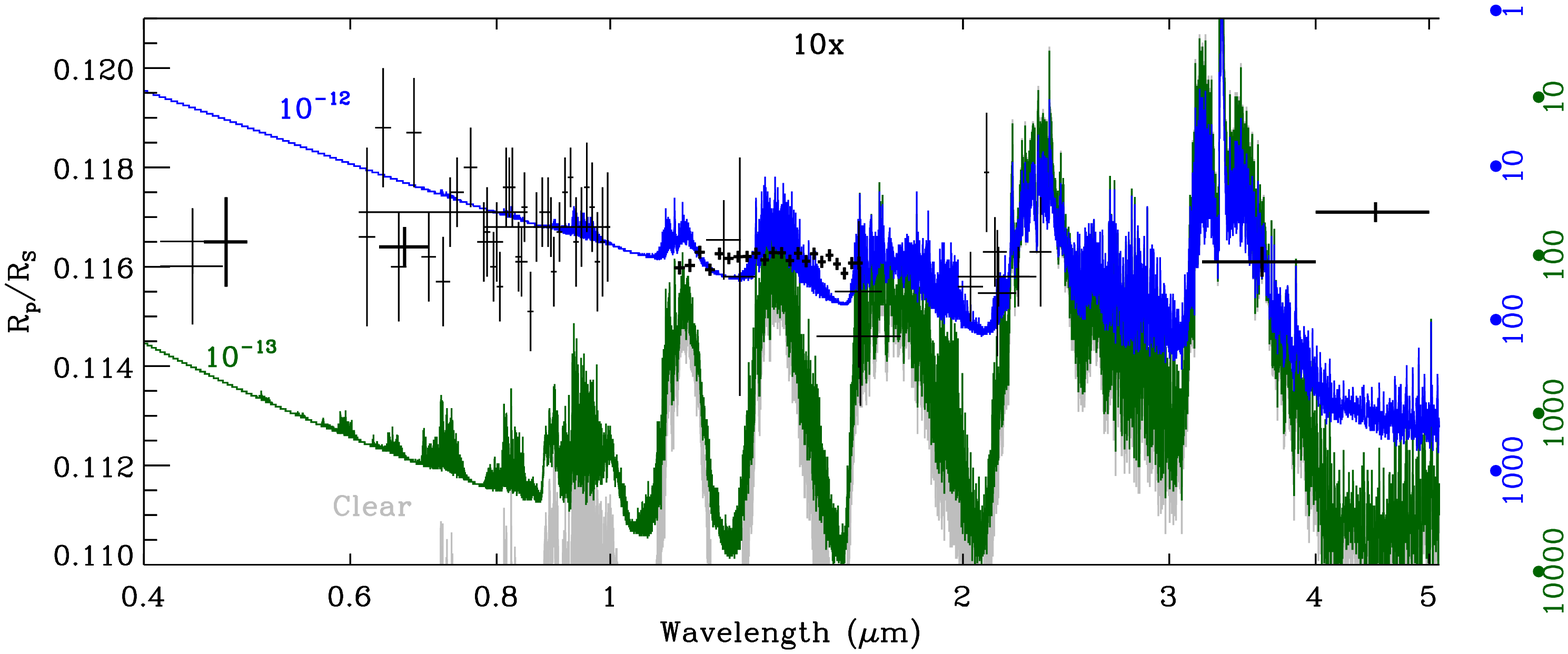}
\includegraphics[scale=0.35]{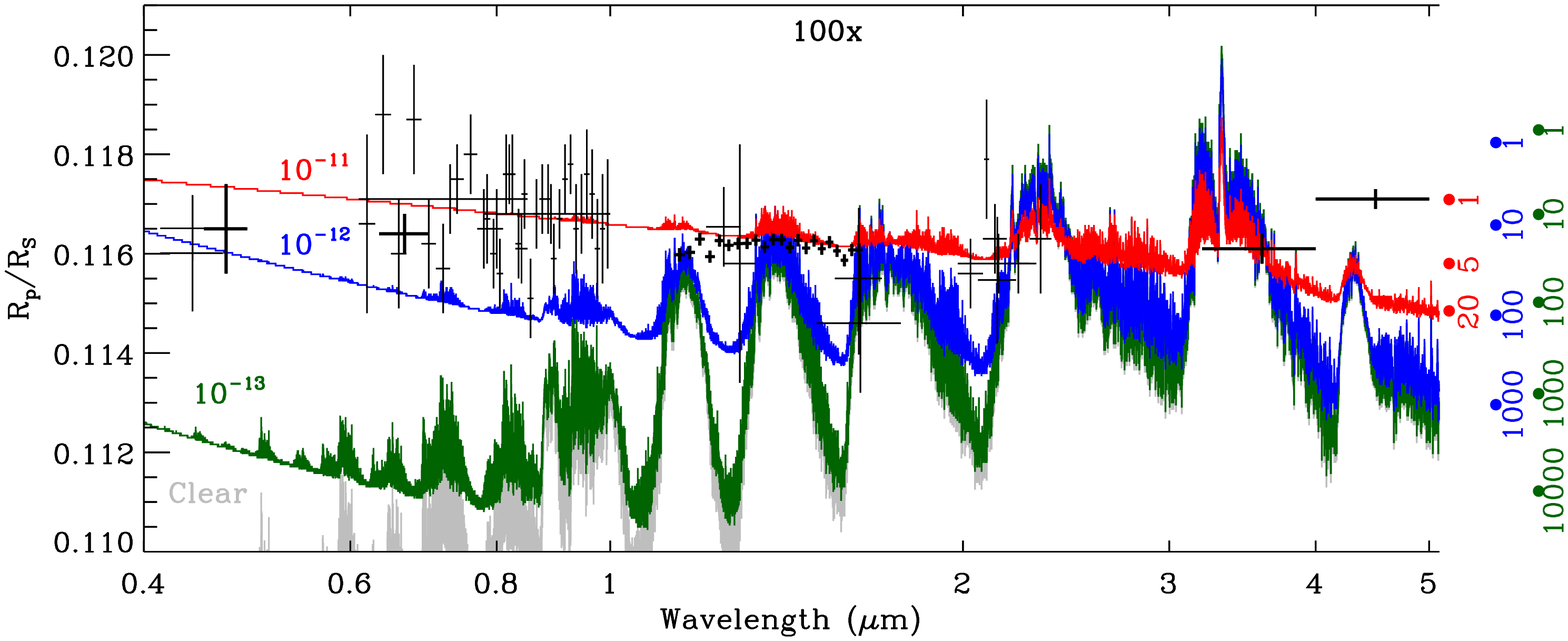}
\includegraphics[scale=0.35]{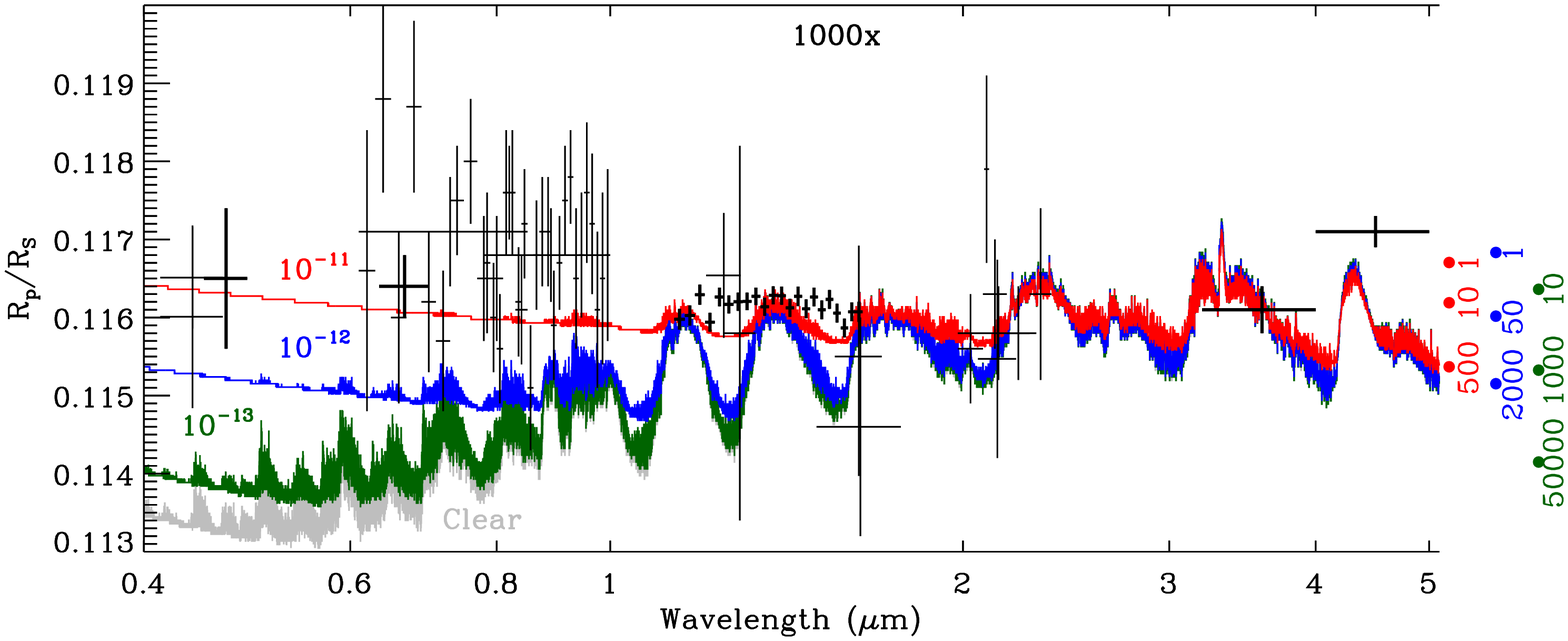}
\includegraphics[scale=0.35]{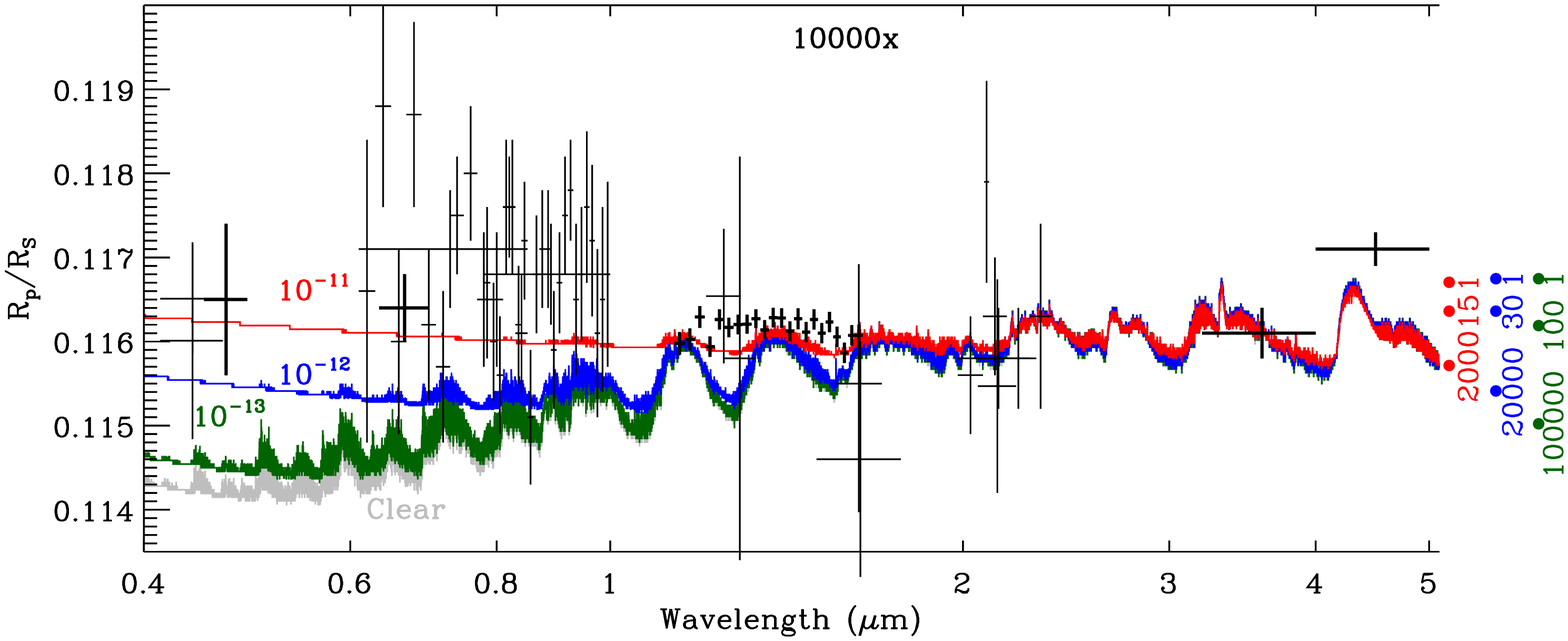}
\caption{Observed (black crosses) and simulated (color lines) transit depth of GJ 1214 b. Each panel presents a different metallicity case and each color line a variation on the assumed haze mass flux. The points and associated numbers on the right of each panel present the pressure (in $\mu$bar) probed at each wavelength. Note that for clarity the pressure range plotted in each panel is different.}\label{transit}
\end{figure*}

Within each metallicity case, increasing the mass flux results in larger particle sizes as already anticipated from our previous study on hot-Jupiters \citep{Lavvas17}. When comparing solar metallicity simulations, the current results are close to those derived for hot-Jupiters at similar conditions of mass flux and atmospheric mixing, as long as, the mixing efficiency is large enough to compensate for the reduction in the characteristic sedimentation velocity of the particles in hot-Jupiters due to their stronger gravitational field. Moreover, due to the cooler thermal structure, particle thermal decomposition is not efficient in the pressure range under investigation for GJ 1214 b.

For these calculations we assumed that particles are not charged thus there is no impedance in their coagulation. However, particle charging will occur in the atmosphere as a result of photoionization or interaction with charge carriers such as electrons and ions. Charge effects become more efficient with increasing particle size as larger particles allow for a higher number of accumulated charges. A detailed treatment of the particles charge distribution is beyond the scope of this evaluation. However, we evaluated the impact of different charge densities, $\chi$, on the particle size distribution  \citep{Lavvas17}. Our calculations suggest that for typical $\chi$ values considered for photochemical hazes (10 - 30e$^{-}$/$\mu$m), particle distributions are affected mainly at pressures greater than 100 mbar with the average particle size reducing by factors between $\sim$1.1 and 1.6 for 10 and 30 e$^{-}$/$\mu$m, respectively, relative to the $\chi$=0 case at solar metallicity (and mass flux of 10$^{-12}$ g cm$^{-2}$s$^{-1}$). However, as the average particle size decreases for higher metallicities, charge effects becomes less important and at 10000$\times$solar metallicity they have no impact on the particle size distribution for the above $\chi$ values. At the pressure range where charge effects are important, haze particles are likely to be lost due to condensation (see Fig.~\ref{temperature} and discussion below), thus the particle size distribution will be drastically different. Therefore, we will not discuss further this effect here.


\section{Transit spectra}
Our simulated transit spectra for different metallicities reveal the impact of the photochemical haze distributions for each case (Fig.~\ref{transit}). Due to the degeneracy between planet radius and corresponding pressure for gaseous planets, we need to assume a reference planetary radius to which all simulations are refered \citep{Lavvas17}. For our simulated spectra we used the 3.6 $\mu$m Spitzer observation as the reference point \citep{Fraine13}, as these observations have small uncertainty and we wish to utilise the measurements at shorter wavelengths for the characterisation of the haze properties. 

The simulated spectra {\color{\clr}of clear atmospheres} demonstrate how the increase in metallicity makes the transit signatures more shallow due to the drop of the atmospheric scale height. However, even for the highest metallicity case considered (10000xsolar) a clear atmosphere (considering only the gas phase abundances) cannot reproduce the observed flatness as the transit depth at visible is always short of the observations \citep{Nascimbeni15}, while the H$_2$O bands provide much deeper troughs than the observations reveal in the near IR \citep{Kreidberg14}. Thus, a heterogeneous opacity is necessary to explain the observations. 

Even if low metallicity cases do not apply to the atmosphere of GJ 1214 b, it is useful to highlight the variation of photochemical haze contribution in the observed spectra for consideration in other planetary cases \citep{Crossfield17}. At low metallicities, we saw that the atmospheric scale height is large and mixing time scales are low. Thus, haze particles remain at high altitudes with a large abundance and affect significantly the observed transit spectrum at visible wavelengths, even for very low formation yields ($\sim$0.01$\%$ for solar metallicity). As metallicity increases, haze opacity in the upper atmosphere decreases and for a given haze mass flux the pressure probed during transit moves to higher pressures, since photons can penetrate to deeper layers before haze opacity sufficiently increases to block them (Fig.~\ref{transit}). For metallicities at and above 100$\times$solar, a higher haze formation yield needs to be considered in order to bring the simulated spectra close to the observations. A mass flux of the order 10$^{-11}$ g cm$^{-2}$ s$^{-1}$ provides enough opacity in the visible/near-IR to bring the transit spectrum close to the observations. Such a mass flux suggests a haze formation yield of $\sim$10$\%$ - 20$\%$. However, as we discuss below, contributions from the photolysis of CO/CO$_2$ could further enhance the haze mass fluxes in which case a common $\sim$10$\%$ for all high metallicity cases (above 100$\times$ solar) is sufficient to reproduce the observations.

The observed  variation of the transit signature among the two Spitzer wavelengths, with the transit depth at 4.5 $\mu$m higher than at 3.6 $\mu$  by $\sim$4$\sigma$ suggests that the higher metallicity end provides the most representative case among those we consider. In our simulations the increase of CO$_2$ with increasing metallicity reduces the high contrast between the two bands characteristic of low metallicities and brings them closer to the observed behaviour at the 10000$\times$solar metallicity case. However, even in this case the simulated ratio of transit at the two wavelength bands is not as close as in the observations suggesting a further enhancement of CO$_2$ relative to our calculations.


\section{Aggregation}

So far we assumed that the haze particles grow as spheres. However, aggregation is a physical consequence of solid particle collisions, and is likely to occur for the photochemical hazes we consider here. The onset of aggregate formation occurs when mass addition from the gas phase is negligible compared to the mass added through particle collisions. Vice versa, aggregated particles can grow to a spherical shape if the rate of mass addition by gas deposition (heterogeneous processes) is larger than the rate of mass addition due to coagulation, as demonstrated by experiments and simulations \citep[see][and references therein]{Lavvas11a}. Such a shape transition is more feasible when particles have a small size, which is why in our previous study of hot-Jupiters we considered only spherical growth. Our simulations suggested that the particle size required to match the transit observations of HD 189733 b was of the order of few nm, at which size heterogeneous reactions on the particle surface could readily preserve the sphericity of the particles \citep{Lavvas17,Lavvas11a}. On the contrary, our simulations for GJ 1214 b, depending on the assumed metallicity, show that larger particle sizes are possible to affect the transit signature. Larger particles in the upper atmosphere are mainly characteristic of low metallicity conditions, which appears inconsistent with the near IR transit signatures. However, aggregate particles have different optical and microphysical properties (in terms of their collisional efficiency and aerodynamic drag) from spherical particles, which may result to a different transit depth signature. 
For example, an aggregate particle has a larger cross section than the same mass spherical equivalent, resulting in higher coagulation rates, hence, faster particle growth. Moreover, aggregates with structures typical of those found in Titan's atmosphere (D$_f$=2, see below) have sedimentation velocities representative of their primary particle size in the free molecular regime \citep{Cabane93,Rannou10,Lavvas10}. Thus, the rapid growth of aggregates will not be limited by an increase of their settling in the upper atmosphere, as would be the case for spherical particles. Therefore, we explore here the haze properties for aggregates.

\begin{figure}[!t]
\centering
\includegraphics[scale=0.5]{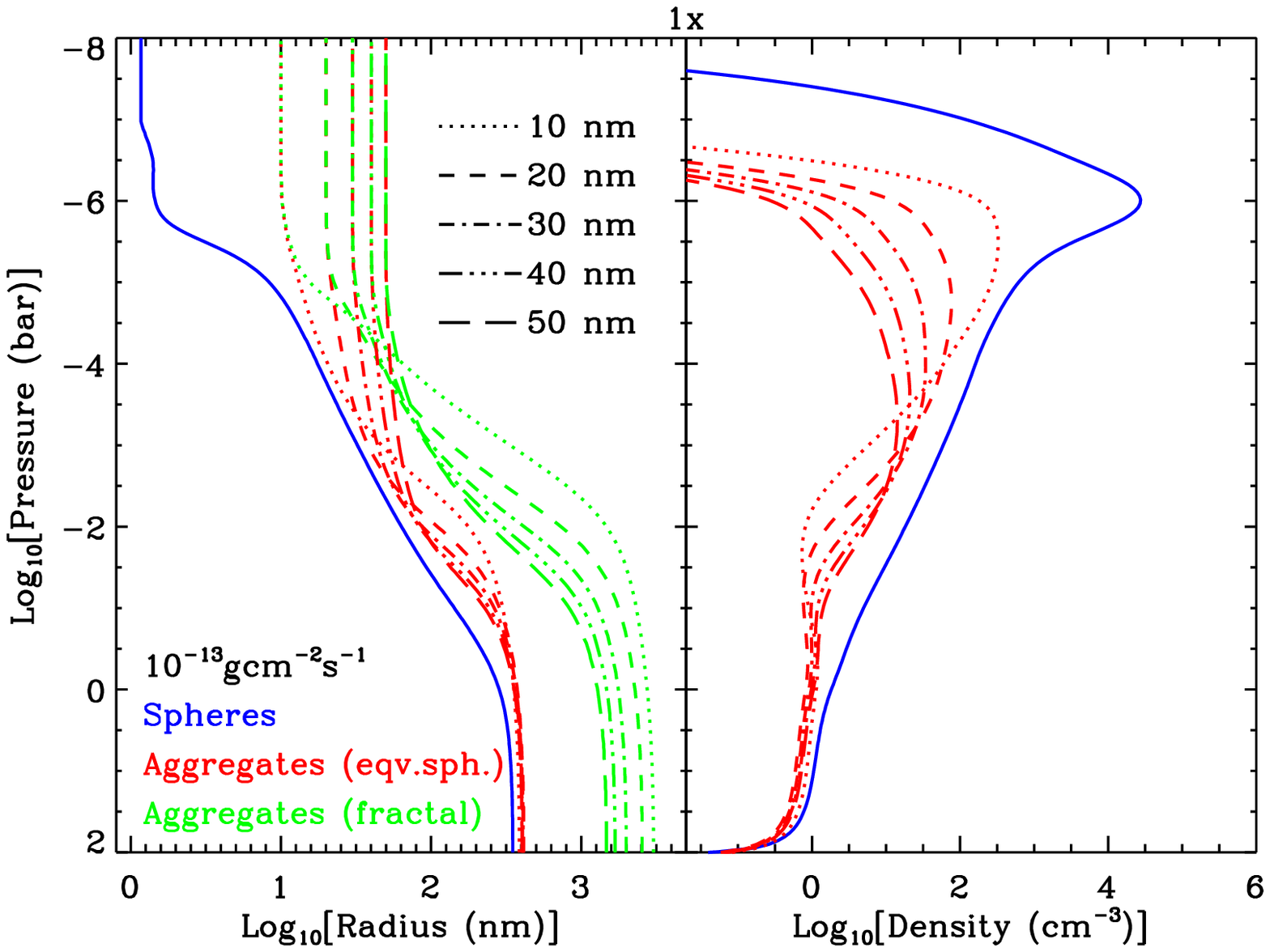}
\includegraphics[scale=0.5]{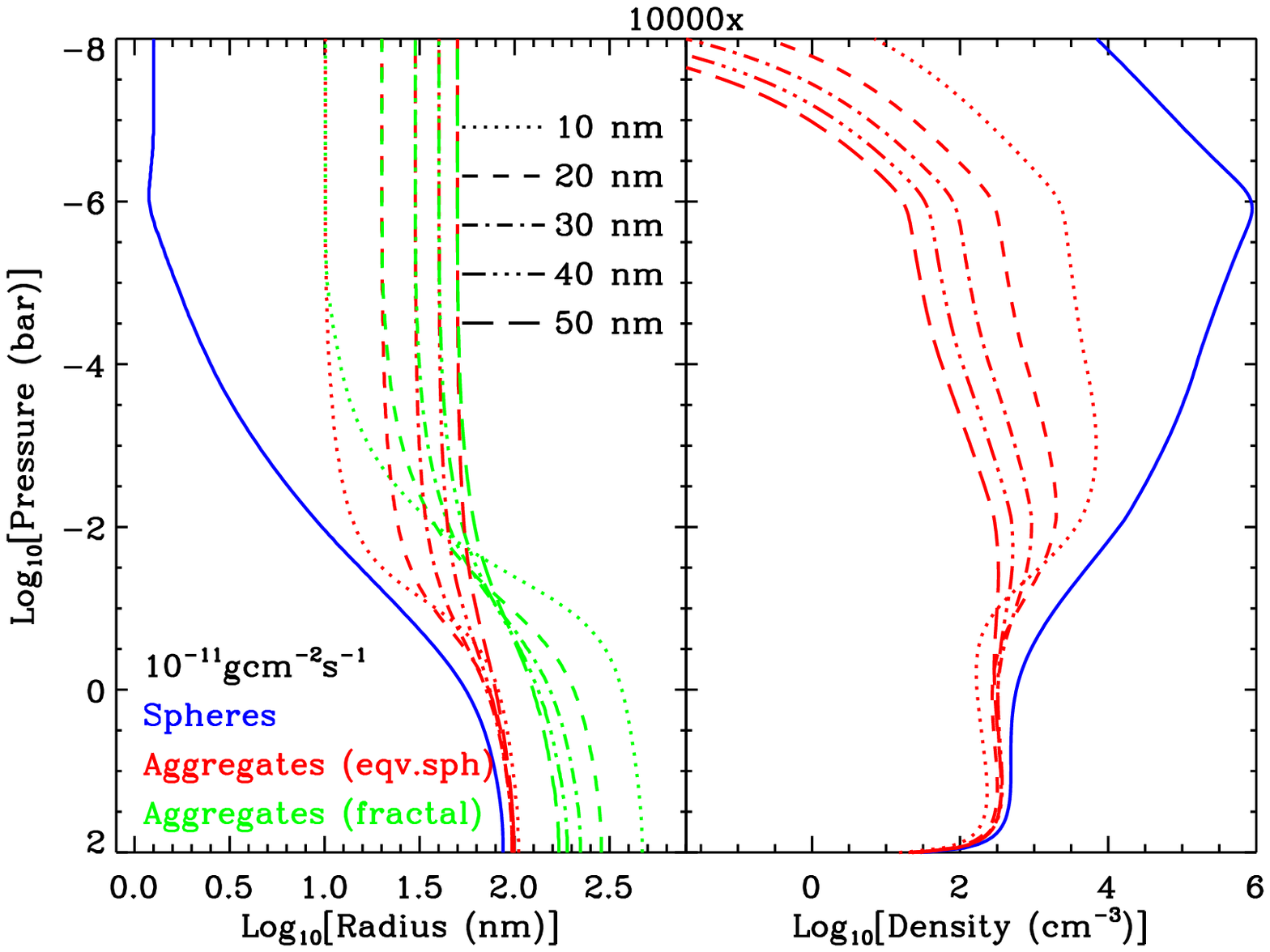}
\caption{Comparison of spherical (blue) and aggregate (red $\&$ green) growth of haze particles for 1$\times$ and 10000$\times$ metallicity cases. Each broken line corresponds to a different primary particle radius. For aggregates the red lines correspond to the average radius of the equivalent mass spherical particle and the green lines to the fractal structure average radius. The corresponding average particle density is shown on the right panels (it does not depend on the radius-type averaging). Note that the assumed mass fluxes for each metallicity case are different and correspond to those providing results close to the observed transit spectra. }\label{aggregates}
\end{figure}

\begin{figure}[!t]
\centering
\includegraphics[scale=0.5]{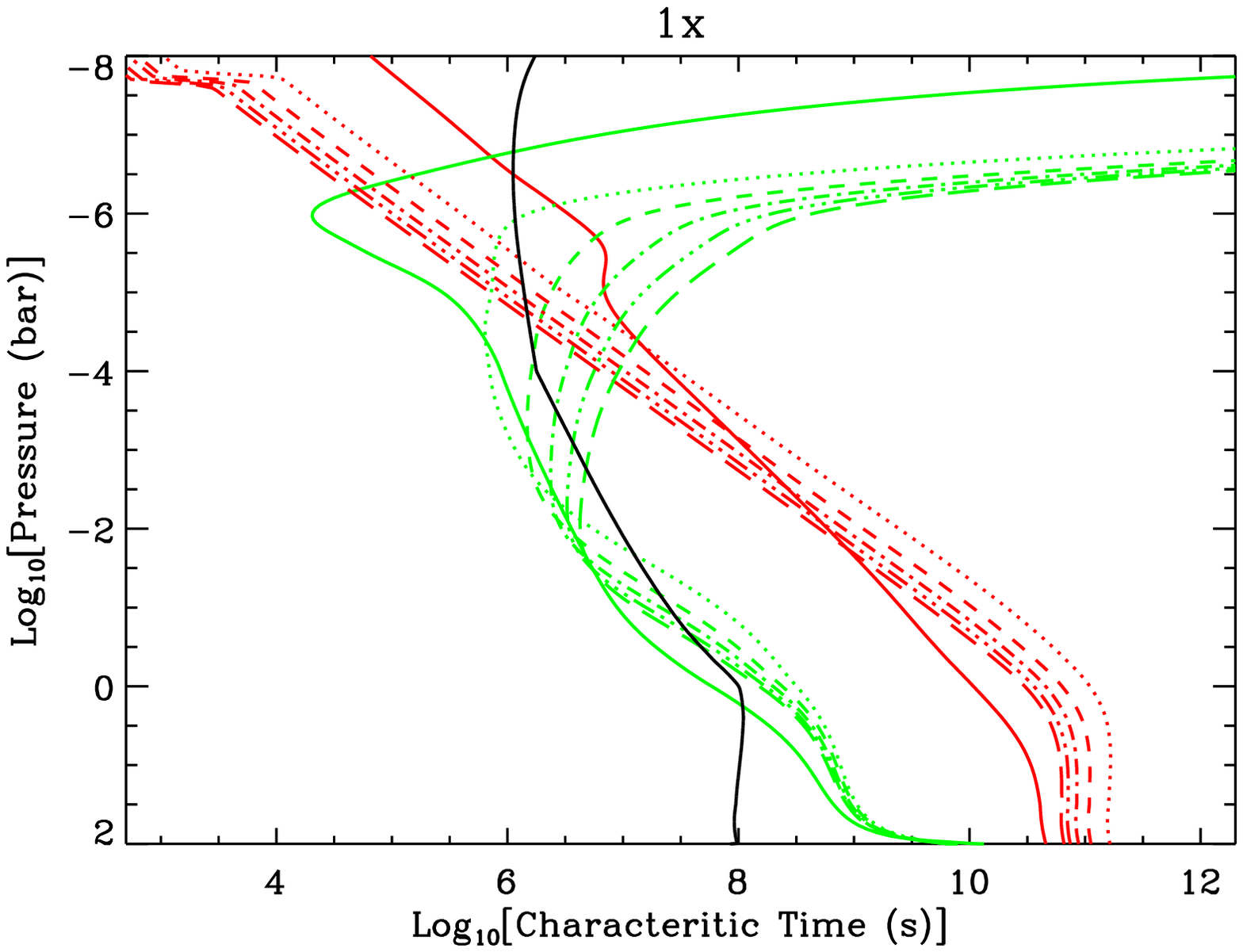}
\includegraphics[scale=0.5]{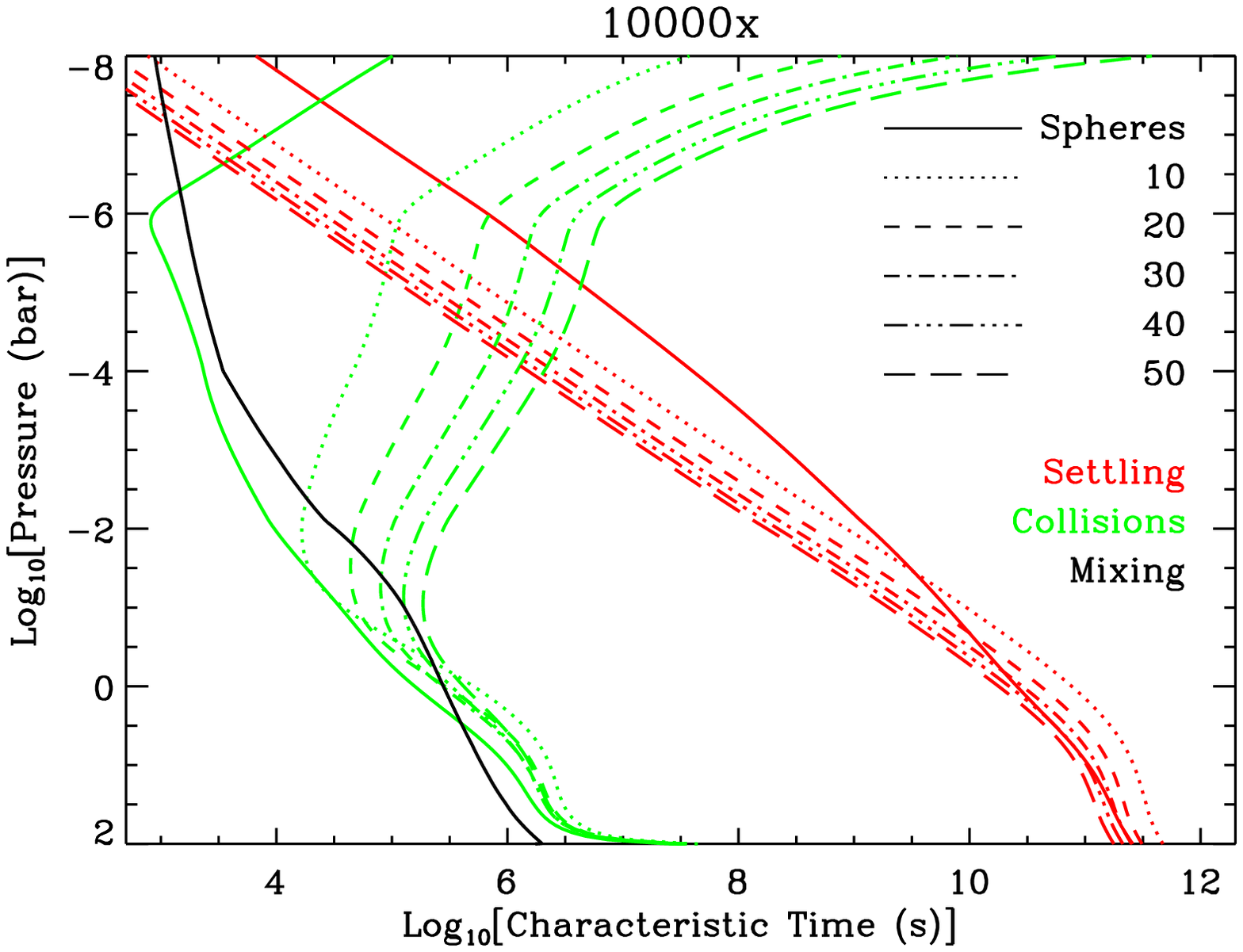}
\caption{Comparison of characteristic times for settling (red), collisions (green), and sedimentation (black) for spheres and aggregates of different primary particle radius (shown in nm).}\label{timesAGR}
\end{figure}

An overview of the properties of aggregate particles can be found in previous studies on Titan's atmosphere, where hazes are known to be of aggregate structure \citep[see review by][and references there in]{West14}. In a nutshell, particle aggregation under atmospheric conditions results in particle structures that are characterised as fractals with a fractal dimension, D$_f$. This means that the distribution of primary particles composing the aggregate structure satisfies a relationship $N_p\sim R_f^{D_f}$, with N$_p$ the number of primary particles and R$_f$ the {\color{\clr}fractal} radius of the aggregate \citep[see][and references therein]{Lavvas10}. The value of D$_f$ depends on the conditions at which the aggregate is formed. Theoretical studies  \citep[see][and references therein]{Cabane93} demonstrate that during ballistic collisions (when the particle mean free path, $\lambda_p$, is much larger than R$_f$) collisions between primary particles and aggregates lead to D$_f$$\sim$3, while collisions between aggregates result in D$_f$$\sim$2. At the opposite extreme of the continuous regime ($\lambda_p <<R_f$) the corresponding fractal dimension limits are D$_f\sim$2.5 and D$_f\sim$1.75 for primary particle-aggregate and aggregate-aggregate collisions, respectively. These pure microphysical cases can be further modified by heterogeneous processes on the surface of the particles, as discussed above, which may result in a further increase of D$_f$. 

The combined contributions of photochemistry that leads to the inception of particles, heterogeneous chemistry on the particle surface, and particle collisions, define the primary particle radius \citep{Lavvas11a}. Typical values of primary particle radii from atmospheres of our solar system are of the order of 10s of nm, e.g. 10 nm for the aggregates in Pluto's atmosphere \citep{Gladstone16} and 40 nm for Titan's haze \citep{Tomasko09}. As we do not yet have a clear picture of this complex mechanism for the conditions under investigation, we treat the primary particle radius as a free parameter and evaluate results for primary particle radii between 10 nm and 50 nm. Moreover we assume that D$_f$=2 below the particle production region. This is a valid approach as below the production region where aggregate-aggregate collisions will dominate, ballistic conditions apply. We estimate that the continuous growth regime applies at pressures higher than $\sim$0.1 bar where $\lambda_p$=0.1$\times$R$_f$ for our calculated size distributions. There, the simulated particles would already be consumed in the formation of clouds, thus modifying D$_f$ will not provide a physically improved solution. \cite{Adams19} assumed a D$_f$=2.4 in their simulations of aggregate growth considering that restructuring of the aggregates due to motion of the primary particles within each aggregate and due to condensation is possible. Although these processes are possible we choose D$_f$=2 in our simulations as this value has reproduced well the haze properties in Titan's atmosphere \citep{Lavvas10, Rannou10}, while a better understanding of the role of condensation on the fractal dimension requires a coupled description with the condensing gases. Further below we discuss the role of hazes in condensation, but we choose first to simulate here the pure haze case and demonstrate the possible differences inflicted by the choice of fractal dimension.





\begin{figure}[!t]
\centering
\includegraphics[scale=0.5]{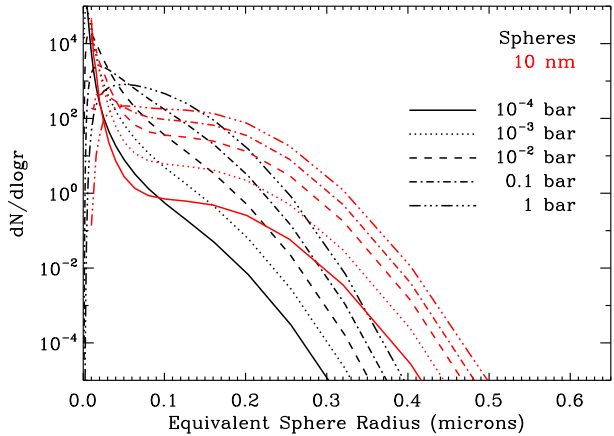}
\caption{Particle size distribution for spherical (black) and aggregate growth (red, 10 nm primary particles) for the 10000$\times$solar metallicity case and for a haze mass flux of 10$^{-11}$gcm$^{-2}$s$^{-1}$. Each line style corresponds to a different pressure level as shown.}\label{distribution}
\end{figure}

Compared to results from the previous simulations of spherical growth where the produced particles were assumed to have a radius of 1 nm, the aggregate simulations provide much smaller particle number densities in the upper atmosphere, as for the same production rate the mass is distributed into larger particles (Fig.~\ref{aggregates}). As the collision rates for aggregates are larger than the corresponding rates for same mass spheres, there is a fast growth of the particles with a corresponding decrease in the particle number density. This is evident when comparing the the average radius during spherical growth (blue lines) with the average radius of the equivalent spherical mass of the aggregate distribution (red lines). The latter is larger than the former due to the porosity of the aggregates that forces the same mass to be distributed over a bigger particle volume. Note that all cases reach a common limiting value of average size and number density in the deep atmosphere. This limit exists because for a given mass flux, collisions limit the particle number density thus once the abundance of particles is sufficiently reduced, the growth terminates. 
However, when considering the collisional and optical cross section of the aggregates we need to evaluate the effective cross section of the fractal structures (green lines). These demonstrate the drastically different effect the aggregation has on the particle properties, with average particle radii being up to an order of magnitude larger than the corresponding particles under spherical growth. 

Although the aggregate growth picture is qualitatively similar to that of spherical growth the involved rates are different (Fig.~\ref{timesAGR}). The particle growth is faster for smaller primary particle radii due to the larger number density of particles available for collisions. For larger primary particles, the sedimentation velocity of the formed particles increases and also becomes a limiting factor for the collisional growth in the upper atmosphere, as particles rapidly sediment to the lower atmosphere. For all cases, aggregation results in larger particle radii with corresponding smaller particle number densities compared to the spherical growth case, for the parts of the atmosphere affecting the transit depth. Moreover, the particle size distribution for aggregates is more broad than the corresponding distribution for spherical particles (Fig.~\ref{distribution}). This effect is a demonstration of the common sedimentation velocity of the particles that makes settling differentiation at a given pressure level less efficient compared to spherical particles. 

\begin{figure}[!t]
\centering
\includegraphics[scale=0.5]{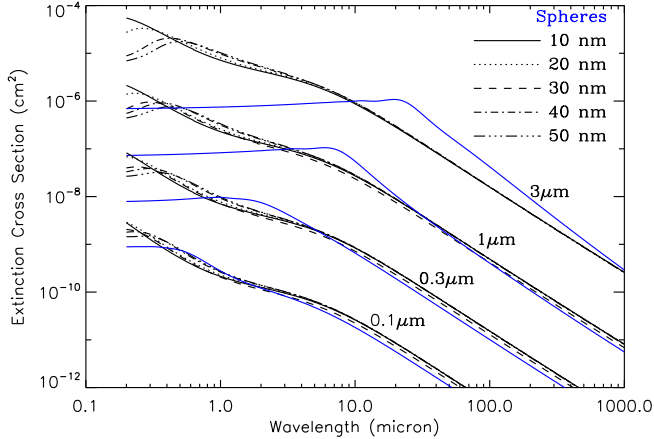}
\caption{Wavelength dependence of the extinction cross section of aggregate particles with different primary particle radii (black lines) compared to the extinction by the same mass spherical particle (blue lines). Four different cases of particles sizes are presented ranging from 0.1 $\mu$m to 3 $\mu$m for the equivalent mass radius.}\label{xsecAGR}
\end{figure}

\begin{figure}[!t]
\centering
\includegraphics[scale=0.5]{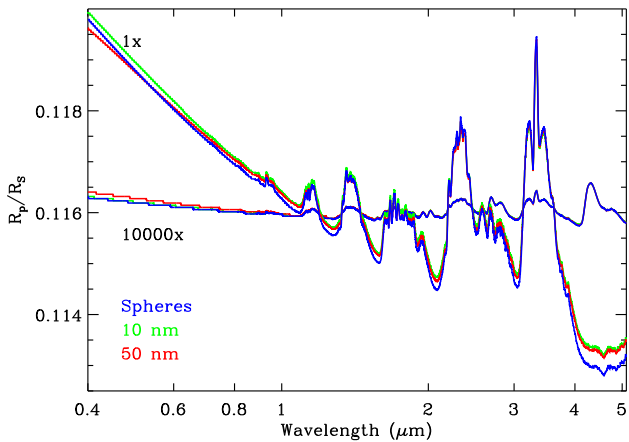}
\caption{Comparison of transit spectra for spherical and aggregate shape haze simulations (see Fig.~\ref{aggregates}) for the 1$\times$ and 10000$\times$solar metallicity cases. Note that spectra are smoothed for clarity.}\label{transAGR}
\end{figure}

 We calculate the optical properties in our aggregate particle distribution utilising the analytical approach based on the T-matrix model results \citep{Tomasko08} assuming a soot composition refractive index \citep{Lavvas17}. The optical properties of an aggregate particle are different from those of the same mass spherical particle and demonstrate higher extinction at UV and lower at IR (see Fig.~\ref{xsecAGR}), with the transition point between these two limits depending on the size and number of primary particles composing the aggregate. A clear characteristic of aggregates is that they result in a sharper wavelength dependence at short wavelengths in their extinction properties, relative to that of equivalent mass spheres. However, when comparing small spherical particles with larger aggregates, i.e. the impact of aggregation on our simulations, although the absolute cross sections are much larger for the aggregates, the wavelength dependence of cross section for the two particle types is similar. Thus, when included in our calculations (Fig.~\ref{transAGR}), we find that aggregate-based transit spectra are very close to those of the spherical particles. Remember that we consider the 3.5 $\mu$m observations as a reference point for the alignment of the pressure scale, \citep[see above and also in ][]{Lavvas17}. Thus, although aggregates provide a higher opacity, their similar spectral signature with smaller spherical particles, provides the same transit signature. This is a different conclusion from what was recently derived by \cite{Adams19}. Based on the results presented by the latter authors, their calculations for the wavelength dependence of aggregate extinction suggest a far more flat wavelength behaviour, similar to what is anticipated for large spherical particles (Fig.~\ref{xsecAGR}). Their calculations are based on a different approach \citep{Rannou97}, which however, provides the same general characteristics for D$_f$=2, with an increase of extinction at UV and decrease at IR relative to the equivalent mass spherical particle. On the other hand \cite{Adams19} assume a fractal dimension of 2.4 in their calculations, which brings the wavelength dependence of the haze opacity closer to that of spherical particles, i.e. more flat \citep[see fig. 3 in][]{Rannou16}\footnote{Note that in fig. 2a of \cite{Rannou16} the assignments of D$_f$ are mistakenly inverted.}, and is the reason for the different conclusion derived by these authors.
 
This similarity between transit spectra for spheres and aggregates demonstrates that extinction measurements can not uniquely constrain the size/shape of particles. In other words, transit observations can be equally well fitted by spherical or aggregate structure particles. Instead polarisation measurements  \citep{Tomasko09, GarciaMunoz18b} or observations at different planetary phases \citep{Seignovert17,GarciaMunoz17} are required for a better characterisation of the particle shape. In our transit depth calculations we did not consider the possible contribution of forward scattering from the particles \citep{Robinson17,GarciaMunoz18}. This contribution will be more important for the aggregate particles that are known to demonstrate a strong forward peak \citep{Lavvas10}. However, recent evaluations of this effect suggest that it reduces the transit depth by less than a scale-height \citep{GarciaMunoz18}. Thus, we did not consider it here.

\section{Discussion}

\begin{figure}
\centering
\includegraphics[scale=0.5]{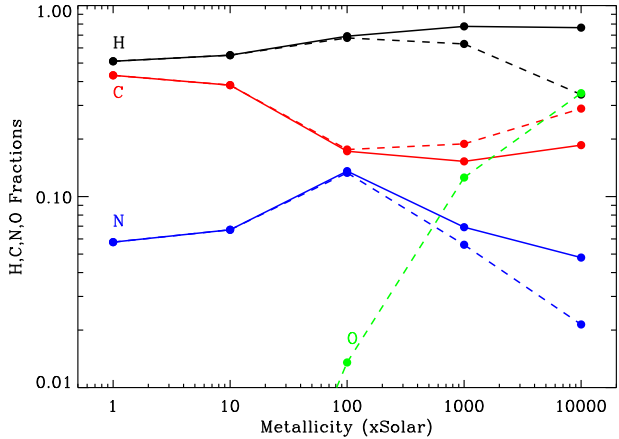}
\caption{Fractional abundance of different elements in the photolysis mass fluxes presented in Table~\ref{Mfluxes}. Solid lines present the case of nominal haze production by photolysis of CH$_4$, N$_2$ and of their main photochemical products, while dashed lines consider the additional contributions from CO and CO$_2$ photolysis.}\label{HCNO}
\end{figure}

Our simulated haze production rates demonstrate that although the metallicity conditions change by 4 orders of magnitude the difference in the haze mass fluxes is only a factor of $\sim$3 between the two extreme cases of solar and 10000$\times$solar metallicity. This is a clear demonstration of the fact that photochemical products are dominantly photon limited, as long as, the abundance of precursors is not critically affected by their photolysis. For example, Pluto and Triton have very similar atmospheres of CH$_4$/N$_2$ composition. Yet, the combination of higher energy input and lower CH$_4$ abundance on Triton relative to Pluto, results in a major destruction of Triton's atmospheric CH$_4$ above $\sim$200 km altitude \citep{Krasnopolsky95} that limits the production of heterogeneous components close to this moons surface, contrary to Pluto where hazes are observed at high altitudes \citep{Gladstone16}. Thus, with the exception of such extreme cases, metallicity changes should not induce major changes to the precursor mass fluxes, thereby, to the haze production rates.

Nevertheless, metallicity changes can impose chemical composition changes with a secondary impact on the particle composition. The relative contributions of H, C, and N in the photolysis of precursors (Fig.~\ref{HCNO}) demonstrate this dependence on the assumed metallicity. Under the nominal soot precursors (i.e. photolysis of CH$_4$, N$_2$ and of their main photochemical products, see above) increasing the metallicity above solar and up to 100$\times$solar, results into a decrease of the available C and an increase of N, while at higher metallicities the C fraction remains approximately constant while N decreases. These fractions only correspond to the composition characteristics of the precursors photolysis mass fluxes and do not necessarily reflect the composition of the soot particles. However, they are indicative of what can be anticipated of the hazes.

\begin{figure}
\centering
\includegraphics[scale=0.5]{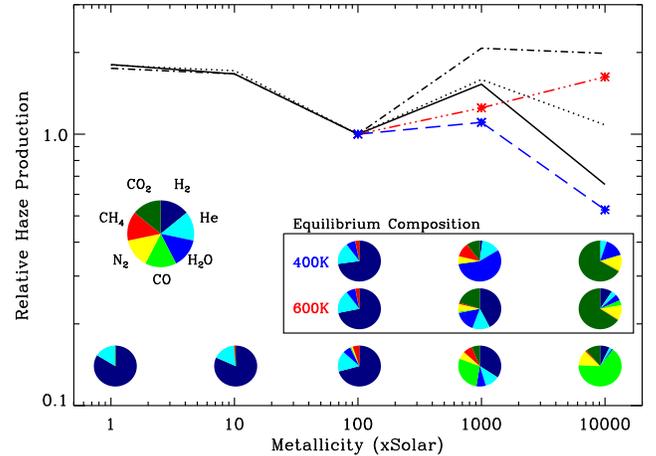}
\caption{Relative haze production from simulations (lines) and laboratory experiments (asterisks) at different metallicities (see text). The black line corresponds to the nominal haze precursors (CH$_4$ $\&$ N$_2$ photochemical products), while the dotted and dash-dotted lines present additional contributions from the photolysis of CO and CO$_2$, respectively. All curves are normalised to the 100$\times$solar metallicity case. Pie charts at the bottom present the main gas composition for each case. }\label{yield}
\end{figure}

The simulated haze production rates at high metallicity conditions are particularly interesting because they can be compared to experimental retrievals of photochemical haze formation at similar conditions. Laboratory investigations for the production of photochemical hazes in super-Earths/mini-Neptunes show production at all metallicity conditions studied (100$\times$,1000$\times$, 10000$\times$), with production rates varying according to the energy source and the temperature conditions \citep{Horst18, He18}. For experiments performed under UV irradiation, which are more relevant to the photochemically produced hazes studied here, particle production for each metallicity case dropped when temperature increased from 400 to 600 K \citep{He18}. For each temperature, particle production among different metallicity cases showed a different dependence (Fig.~\ref{yield}): at 400 K the relative haze production was slightly higher at 1000$\times$ compared to 100$\times$ solar metallicity, and dropped by a factor of 2 at 10000$\times$, while at 600 K a monotonic increase was observed with increasing metallicity. 

A direct comparison with these experimental results is not straight forward as we can compare only relative yields among the different metallicity cases considered. Moreover, our simulations are representative of the photochemistry occurring at 500 K in the upper atmosphere of GJ 1214 b, where the main atmospheric composition is defined by disequilibrium chemistry (Table~\ref{comp1mbar}). On the contrary, the experiments performed at the bracketing 400 K and 600 K temperature cases had a main gas composition based on thermochemical equilibrium conditions at 1 mbar pressure \citep{He18}. Nevertheless, our calculated composition at 100$\times$ and 1000$\times$ solar metallicity cases are not very different from those of the experiments (see pie charts in Fig.~\ref{yield}). At 100$\times$ the disequilibrium CH$_4$ mixing ratio at 500 K (4.6$\%$) is very close to the experimental value used (4.5$\%$ at 400 K and 3.4$\%$ at 600 K), and the rest of the main composition is similar between our calculations and the experiments with only difference a small contribution ($\sim$1$\%$) of N$_2$ in our calculations. At 1000$\times$ our CH$_4$ mixing ratio (6.7$\%$) is in between the two equilibrium abundances (11$\%$ at 400 K and 1.7$\%$ at 600 K). However the main gas composition in our simulation is based on H$_2$ (34$\%$) and CO (27$\%$), while for the experiments the main equilibrium composition changes from H$_2$O (56$\%$) at 400 K to H$_2$ (42$\%$) and CO$_2$ (20$\%$) at 600 K. These changes in the main composition of our simulations occur due to quenching in the lower atmosphere by the atmospheric mixing (Fig.~\ref{composition}). However the relative contributions of C-H/C-O species are qualitatively similar between our calculations (6.7/33.8) and the experiments (11/10 at 400K and 1.7/21.9 at 600 K). This picture drastically changes at 10000$\times$solar metallicity case for which the disequilibrium composition includes $\sim$1$\%$ CH$_4$, while for the equilibrium conditions methane is a minor species not included in the experimental gas mixture. In addition, the main composition in the disequilibrium conditions is based on CO (65$\%$) with contributions from CO$_2$ (11$\%$) and N$_2$ (12$\%$), while at equilibrium CO$_2$ (67$\%$) is the main gas component (Fig.~\ref{yield}). There are also differences in the spectral density of the UV light used in the experiment and that of GJ 1214, that impose further reasons for differentiation in the results. Ly-$\alpha$ (121.6 nm) dominates the UV stellar spectrum, while the laboratory UV lamp has a stronger output near 160 and 230 nm relative to 121.6 nm. Notwithstanding these differences between experimental and theoretical derivations we can identify some qualitative similarities. 

For both experiments and theoretical calculations increasing the metallicity from 100$\times$ to 1000$\times$solar results in higher haze production (Fig.~\ref{yield}). Although the photolysis mass fluxes from the disequilibrium chemistry suggest a stronger increase than the experimental results at 400 and 600 K, qualitatively both theory and experiments provide the same picture, and the observed differences are likely related to the differences in the gas abundances and UV spectral density identified above. For example, at 1000$\times$solar metallicity H$_2$O and CO$_2$ have significant abundances in the experimental mixtures at 400 and 600 K respectively, which are replaced by H$_2$ and CO in our calculations. The latter two gases do not intervene in the photolysis of hydrocarbons contrary to the former two gases that have major absorption bands at UV wavelengths and will screen the photolysis of methane and other hydrocarbons resulting in smaller haze production. 

At 10000$\times$solar metallicity, our calculations show a clear decrease in the haze production that is consistent with the experimental results at 400 K. However, our estimated haze production is based only on hydrocarbons and N-containing species (N$_2$, NH$_3$, HCN), while the experimental conditions do not include any CH$_4$. This characteristic suggests that formation of photochemical hazes from the photolysis of CO/CO$_2$ is also possible \citep{He18, Horst18}. If we consider the possible contribution of carbon from the photolysis of CO in the theoretical haze mass flux estimates (dotted curve in Fig.~\ref{yield}) we derive a higher haze production, which becomes even higher if we also consider the photolysis of CO$_2$ (dash-dotted curve). In addition, the fractional C-abundance increases with these contributions at high metallicity cases, suggesting that the hazes become more carbonaceous at these conditions (Fig.~\ref{HCNO}). 

The CO/CO$_2$ contributions bring the relative haze production close to the metallicity behaviour seen at 600 K, but they also increase further the production at the 1000$\times$solar metallicity case.  A naive interpretation of the experimental results at the two temperatures would suggest that the relative contribution of CO/CO$_2$-produced hazes is higher at 600 K than at 400 K. Although such a behaviour may partially contribute to the observed behaviour, these relative yields have to be examined through the prism of the main gas composition. The main difference of the gas composition at the two temperatures studied in the laboratory experiments at 10000$\times$solar metallicity, is that at 400 K there is $\sim$3$\times$ more H$_2$O than at the equilibrium composition at 600 K, that could impose a significant reduction on the haze formation, again through photolysis screening of other species, as well as, by oxidizing reactions. In our corresponding disequilibrium simulations the H$_2$O abundance (1.2$\%$) is closer to the 600 K case, for which we get a similar relative production. 

In addition, differences in the chemical rates imposed by the different temperature conditions could further affect the haze production. We note that experimental and theoretical results correspond to the different steps of the photochemical haze production: the experiments measure the abundance of the formed haze particles i.e. the end products of the photochemistry, while our theoretical estimates are based on the first steps of the atmospheric photochemistry. These two extremes are related by the haze formation yield. The qualitative consistency between our theoretical simulations and the experimental results verify the appropriateness of this approach. However, the comparison also demonstrates that utilising the photolysis mass fluxes provides a more robust estimation of the haze production, instead of scalings based on the abundances of individual precursors. This is readily demonstrated by observing that abundances across metallicity cases demonstrate a monotonic variation (e.g. CH$_4$ and HCN in Fig.~\ref{composition}) that would result in monotonic haze production yields across metallicities, inconsistent with the experimental results. 

\begin{figure}
\centering
\includegraphics[scale=0.5]{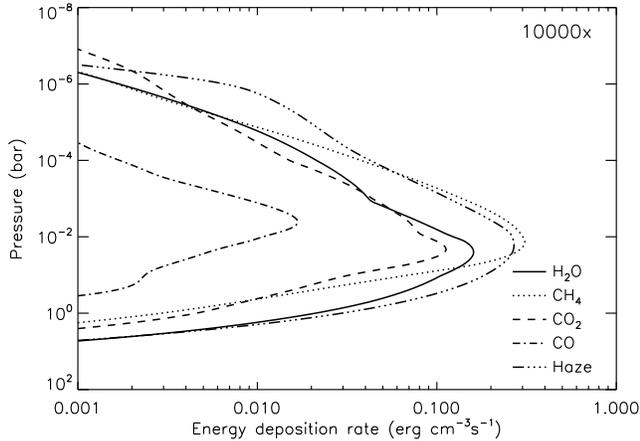}
\caption{Energy deposition in the atmosphere of GJ 1214 b from the absorption of stellar radiation. Here we compare the haze contribution with the main gaseous absorbers. }\label{heating}
\end{figure}

We highlight that the above discussion applies only to the relative production yields. The experiments reveal that absolute yields increase with decreasing temperature for the 100$\times$ and 1000$\times$ metallicity cases, while the opposite behaviour was observed for the 10000$\times$solar metallicity case \citep{He18}. The absolute haze production rates for the latter case were measured to raise from 0.01 mg hr$^{-1}$ at 400 K to 0.013 mg hr$^{-1}$ at 600 K. The ratio of 1.3 between these to values is tantalisingly close to the value of 1.35 we estimate for the ratio of photolysis rates, J$_{\rm 600K}$/J$_{\rm 400K}$, of a pure CO$_2$ gas at these two temperature limits \citep{Venot18}, assuming the spectral density of the UV lamp (Fig.~\ref{star}). Given that CO$_2$ is the main gas component at this metallicity, it is logical to correlate the temperature dependence of the CO$_2$ cross section with the observed variation of the haze production. At lower metallicities, the contribution of CO$_2$ is smaller therefore its impact on the haze production is smaller. However, the reason for the inversion of the haze yield with temperature is not clear and has to be sought in the chemical pathways leading to the haze formation. Such an investigation necessitates a detailed comparison with the experiments at identical conditions (gas mixtures/ energy input) to identify the role of each process on haze production, and which will further benefit from comparison with the experimentally derived gas phase products \citep{He19}. Such a study goes beyond our goals for the current investigation. However, the conclusion we derive from this comparison of relative production yields is that the theoretical simulations capture the qualitative behaviour identified in the experiments and vice versa the experiments provide a picture of haze formation representative of theoretical anticipations, indicating promising results from this venue of investigation. 


\begin{figure}[!t]
\centering
\includegraphics[scale=0.5]{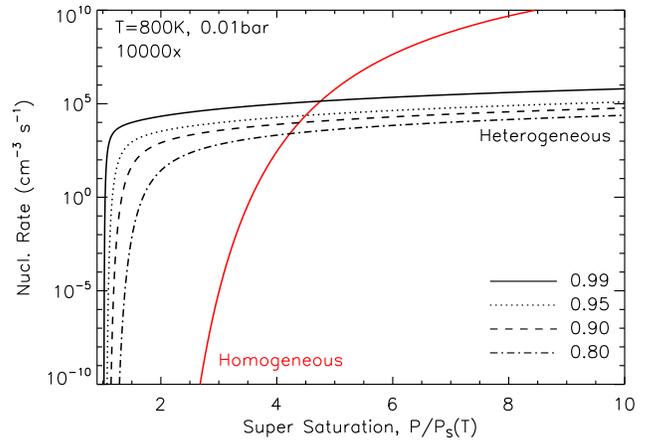}
\caption{Homogeneous (red) and heterogeneous nucleation rates for KCl. For the heterogeneous rates we consider our simulated haze particle distribution (spheres) for the 10000$\times$solar metallicity case at 800 K and 0.01 bar. Different curves correspond to various values for the contact angle, $\theta$, between the substrate (haze particle) and the condensed phase, expresses as $\mu$=cos($\theta$).}\label{nucleation}
\end{figure}

We now turn our attention to the implications of the haze particles on the atmosphere. We find that our simulated haze distributions (of either spherical or aggregate shape) do not impose major modifications to the radiation field at UV wavelength, therefore do not have a major impact on the simulated chemical abundances. This occurs because for the sub-Neptune conditions investigated here gases provide a larger UV opacity than haze, under all metallicity conditions. For example, at solar metallicity conditions for which we can compare with previous evaluations for hot-Jupiters \citep{Lavvas17}, H$_2$O, NH$_3$, and CH$_4$ have higher abundances for sub-Neptune conditions, therefore dominate the UV opacity. 

On the contrary, at longer wavelengths where gas absorption decreases, haze opacity dominates. Thus, as with the case of hot-Jupiter hazes, we find that the presence of such particles in the atmosphere could significantly affect the thermal structure. Our calculations show that absorption of stellar radiation by the particles results in the same energy deposition rate as for the main gas absorbers. For the 10000$\times$solar metallicity that is closer to the observed spectra, CH$_4$ is the dominant gas absorber followed by H$_2$O. Our calculations suggest that the haze contribution in the energy deposition is similar to that of CH$_4$ and {\color{\clr}at pressures lower than} 0.1 mbar dominates over all gas contributions up to 1 $\mu$bar (Fig.~\ref{heating}). Previous studies, focusing on the impact of cloud particles on the energy balance found that they could significantly affect the atmospheric thermal structure \citep{Charnay15b, Roman19}. Soot composition particles are expected to have a larger impact as they demonstrate higher absorptivity than the cloud composition components. However, the extent of this impact will depend on how efficiently haze particles are consumed as condensation nuclei in cloud formation. 

Our preliminary estimates for the impact of nucleation demonstrate that haze particles would efficiently act as nucleation sites. Our calculations for the condensation of KCl that is expected to condense at higher altitudes compared to ZnS \citep[see Fig.~\ref{temperature} and][]{Gao18}, demonstrate haze particles would readily lead to the formation of cloud particles at very low super-saturations (Fig.~\ref{nucleation}). Even for small values of the contact angle between the haze particle and the condensed phase formed on its surface, a parameter that is currently unknown, the heterogenous nucleation rate dominates over
the homogeneous nucleation rate at low super-saturations. Currently in the simulations of cloud formation in mini-Neptunes/super-Earths \citep{Ohno18, Gao18} KCl condensation is assumed to start through homogeneous nucleation, while ZnS is formed heterogeneously on the surface of the KCl cloud particles. Thus, the different population of nucleation sites imposed by the presence of the photochemical hazes will have implications for the properties of all subsequent cloud particle distributions. Therefore, our results demonstrate that a coupled picture between photochemical hazes and cloud formation is necessary. 

{\color{\clr}Our simulations provide a disk-average picture of haze properties for the atmosphere of GJ 1214 b, which will be further modulated by the atmospheric circulation. Although the assumed K$_{ZZ}$ profile in our study does take into consideration the impact of advection on the vertical direction, horizontal re-distribution of the hazes may further modify their impact on the energy deposition, as well as, their transit signature \citep{Line16}. Previous studies have mainly focused on the impact of temperature changes on the properties of clouds, but studies with passive tracers in GCM (more consistent of what should be anticipated for haze particles) demonstrate that circulation can result in significant inhomogeneity in the particle distribution across the planetary terminator \citep{Parmentier13, Charnay15a}. Moreover, horizontal modifications of the chemical composition may further affect the local haze production, which we assumed here to be dominated by the dayside production rate. Finally, modification of the atmospheric properties along the line of sight of transit observations can modify the observed planet radius \citep{Caldas19}. Although such an effect appears to be smaller for sub-Neptune planets relative to hot-Jupiters it has to be evaluated for each case. Proper evaluation of these processes requires a detailed treatment of the atmospheric 3D properties with a coupled model of circulation, photochemistry $\&$ microphysics. Our present work can serve as a starting point for such an endeavor.}




\section{Conclusions}
Our study reveals the properties of photochemical hazes in atmospheres of super-Earths and mini-Neptunes atmospheres, focusing on the case of GJ 1214 b. We find that photochemical haze formation with a formation efficiency of 10$\%$ relative to the photolytically produced mass fluxes provides haze particle distributions consistent with the observed transit spectrum when high metallicity conditions are considered (10000$\times$solar). We discuss the properties of both spherical and aggregate particles and demonstrate that the two growth mechanisms lead to different particle size distributions, but to similar transit signatures when the fractal aggregate dimension is 2. We also find that the presence of photochemical hazes can have major ramifications for the properties of the condensation cloud particles and that a coupled description including both heterogeneous components should provide a more realistic picture for the properties of heterogenous compounds in such atmospheres as well as for their implications on the thermal structure. 

In a broader sense we find that increasing the atmospheric metallicity can result in non-monotonic changes in the haze production rate, although the difference in the estimated rates are small (factor of $\sim$3) relative to the range of metallicity cases considered (1$\times$-10000$\times$solar). To a first degree explored here we find that these modifications depend on the complex interplay between the dominant gaseous abundances for each metallicity case and their interaction with the incoming stellar radiation. Comparison of our simulations with laboratory experiments of photochemical haze production reveals a qualitatively consistent picture. On a second degree further variations will be imposed by changes on the chemical pathways leading to haze formation that are currently unknown, and which will define the absolute haze formation yield. These are aspects that need to be investigated in the future through combined theoretical and experimental approaches under well constrained conditions. 

\acknowledgments
We thank Prof. P. Rannou for his comments on the impact of D$_f$ on the aggregate optical properties. This work was supported by the Programme National de Plan\'etologie (PNP) of CNRS/INSU, co-funded by CNES. M.S. was supported by NASA Headquarters under the NASA Earth and Space Science Fellowship Program - Grant 80NSSC18K1248. AGM acknowledges the support of the DFG priority program SPP 1992 "Exploring the Diversity of Extrasolar Planets (GA 2557/1-1).

\bibliographystyle{aasjournal}
\bibliography{refs}



\end{document}